\documentclass[a4paper,fleqn,usenatbib]{mnras}

\usepackage{natbib}
\usepackage{color}
\usepackage{pdflscape}

\usepackage[T1]{fontenc}
\usepackage{ae,aecompl}

\usepackage{graphicx}	
\usepackage{amsmath}	
\usepackage{amssymb}	

\newcommand{\flyc}{\textit{NB359}\,}
\newcommand{\flya}{\textit{NB497}\,}
\newcommand{\oxyrat}{[O{\sc iii}]/[O{\sc ii}]\,}
\newcommand{\flycfuv}{$f_{\mathrm{LyC}}/f_{\mathrm{UV}}$}

\title[Subaru $z>3$ LyC search in GOODS-N]{Subaru narrow-band imaging
search for Lyman continuum from galaxies at $z>3$ in the GOODS-N field
\thanks{Based on data collected at Subaru
Telescope, which is operated by the National Astronomical Observatory of
Japan. Proposal IDs: S11A-016, S15A-019, S18A-042.}}

\author[I. Iwata et al.]{
Ikuru Iwata,$^{1,2,3}$\thanks{E-mail: ikuru.iwata@nao.ac.jp}
Akio K. Inoue,$^{4,5,6}$
Genoveva Micheva,$^{1,7}$
Yuichi Matsuda,$^{2,8}$
\newauthor
and Toru Yamada$^{9}$
\\
$^{1}$Subaru Telescope, National Astronomical Observatory of Japan, 650 North A'ohoku Place, Hilo, Hawaii 96720, USA\\
$^{2}$Department of Astronomical Science, The Graduate University for Advanced Studies (Sokendai), 2-21-1, Osawa, Mitaka, Tokyo\\
181-8588, Japan\\
$^{3}$Department of Astronomy and Physics and Institute for Computational Astrophysics, Saint Mary's University, 923 Robie Street, \\
Halifax, Nova Scotia B3H 3C3, Canada\\
$^{4}$Department of Environmental Science, Faculty of Design Technology, Osaka Sangyo University, 3-1-1, Nakagaito, Daito, Osaka \\
574-8530, Japan\\
$^5$Department of Physics, School of Advanced Science and Engineering, Waseda University, 3-4-1, Okubo, Shinjuku, Tokyo 169-8555,\\ Japan\\
$^6$Waseda Research Institute for Science and Engineering, 3-4-1, Okubo, Shinjuku, Tokyo 169-8555, Japan\\
$^{7}$Leibniz Institut f\"ur Astrophysik, An der Sternwarte 16, D-14482 Potsdam, Germany\\
$^8$National Astronomical Observatory of Japan, 2-21-1, Osawa, Mitaka, Tokyo 181-8588, Japan\\
$^9$Institute of Space and Astronautical Science, Japan Aerospace
Exploration Agency, 3-1-1 Yoshinodai, Chuo-ku, Sagamihara City, \\
Kanagawa 252-5210, Japan
}

\date{Accepted 2019 July 22. Received 2019 July 17; in original form 2019 January 7}

\pubyear{2019}

\begin{document}
\label{firstpage}
\pagerange{\pageref{firstpage}--\pageref{lastpage}}
\maketitle

\begin{abstract}
We report results of a search for galaxies at $z>3$ with Lyman continuum
 (LyC) emission using a narrow-band filter \flyc with Subaru /
 Suprime-Cam in a $\sim$800 arcmin$^2$ blank field around the
 GOODS-N. We use 103 star-forming galaxies (SFGs) and 8 AGNs with
 spectroscopic redshifts in a range between 3.06 and 3.5, and 157
 photometrically selected $z=3.1$ Lyman $\alpha$ emitter (LAE)
 candidates as the targets.
 After removing galaxies spectroscopically confirmed to be contaminated
 by foreground sources, we found two SFGs and one AGN as candidate LyC
 emitting sources among 
 the targets with spectroscopic redshifts.
 Among LAE candidates, five sources are detected in the \flyc
 image, and three among them  may be contaminated by
 foreground sources.
 We compare the sample galaxies in the GOODS-N with those in the SSA22,
 where a prominent protocluster at $z=3.1$ is known and the LyC search
 using the same \flyc filter has been made.
 Frequency of galaxies with
 LyC leakage in the SSA22 field may be about two times higher than the
 galaxies in the GOODS-N with the sample UV magnitude range, although
 the numbers of LyC detections in these fields are too small to make a
 statistically significant conclusion.
 By combining the sample galaxies in these fields, we place the
 3$\sigma$ upper limits of the observed LyC-to-UV flux density ratio and
 LyC escape fraction for galaxies at $z=3.1$ with absolute UV magnitude
 $M_\mathrm{UV} < -18.8$ as (\flycfuv)$_\mathrm{obs} < 0.036$ and
 $f_\mathrm{esc}^\mathrm{abs}<8$\%, respectively.
\end{abstract}




\section{Introduction}

A significant portion of photons from young massive stars has wavelengths
shorter than the hydrogen Lyman limit (912\AA) and they can ionize neutral
hydrogen. While some of the ionizing photons emitted from
star-forming regions in a galaxy would be absorbed by dust and
interstellar H{\sc i} gas, some other ionizing photons may escape into
the intergalactic medium (IGM).
The escape fraction of Lyman continuum (LyC), $f_\mathrm{esc}$, is the
fraction of ionizing photons which leak into the IGM among those
produced.
Star-forming galaxies are thought to be primary sources of LyC which
have caused cosmic reionization at $z>7$
\citep[e.g.,][]{Stiavelli2004, Oesch2009}, and
$f_\mathrm{esc}$ is one of the most important parameters necessary to
understand the process of cosmic reionization. However, since LyC is
easily absorbed by the intervening H{\sc i} gas, the direct measurement
of LyC from galaxies at $z\gtrsim5$ is practically impossible
due to high opacity of the IGM. On the
other hand, there is still a significant probability of fairly
transparent lines of sight towards galaxies at $3<z<5$
\citep{Inoue2008}, and we have chances to explore the LyC
escape from galaxies in the young Universe.

Although any detection of LyC from faint high-z galaxies is
observationally challenging, significant observational efforts 
on direct detections of LyC
radiation from distant galaxies have been made.
There are several spectroscopic detections of LyC for individual
galaxies at $z>2$ so far
\citep{Vanzella2010b, Vanzella2015, Shapley2016, Vanzella2018}
and a recent report of a systematic spectroscopic survey of
LyC from 124 Lyman Break Galaxies (LBGs) by \citet{Steidel2018}
at $2.7 < z < 3.6$ includes
15 individual LyC detections. \citet{Steidel2018} also measured the
ratio of observed ionizing to non-ionizing UV flux density \flycfuv\, with
composite spectra, and with correction for the IGM and circum-galactic medium
(CGM) via a Monte Carlo analysis. Through population synthesis
modelling of the intrinsic spectrum of the observed composite spectra,
they constrained the intrinsic $f_\mathrm{esc}$ for their sample
galaxies. the averaged value of
$f_\mathrm{esc}$ for the 124 galaxies with
$-22.1 \leq M_\mathrm{UV} \leq -19.5$ is $0.09 \pm 0.01$.
\citet{Marchi2018} used a set of 201 rest-frame UV spectra
of the galaxies from VUDS \citep{LeFevre2015} at $3.5 < z < 4.3$ to
examine their LyC escape fraction. They constructed subsamples based on
Ly$\alpha$ equivalent widths (EWs), Ly$\alpha$ velocity offsets from
systemic redshift, spatial extents in Ly$\alpha$ emission and rest-frame
UV continuum, and stacked their spectra to constrain LyC/UV flux density
ratios. They found that subsamples with large Ly$\alpha$ EW and those
with smaller spatial extents either in Ly$\alpha$ or in rest-frame UV
have higher relative LyC flux compared to the other galaxies in their
sample, and suggested that these properties would be good indicators of
galaxies with strong LyC emission.

An alternative approach to constrain LyC leakage from galaxies is to use
photometry using a filter with bandpass free from non-ionizing photons.
Because absorption by neutral hydrogen in the CGM and IGM is expected to be
smaller for a wavelength range just below the Lyman limit compared to
that in shorter wavelengths, use of imaging with a narrow-band filter
tracing such a wavelength range for a target redshift could be an
efficient way to search for LyC emitting galaxies in a field.
This narrow-band approach has been employed by several past researches
\citep{Inoue2005, Iwata2009, Nestor2011, Mostardi2013, Micheva2017},
and these studies reported detections of a significant
number of candidates of LyC emitting
galaxies. However, the use of imaging data for LyC search runs the risk of
misidentifying foreground sources as LyC emitters due to chance
overlap. Indeed, follow-up spectroscopic observations of photometrically 
selected candidates of LyC emitting galaxies have revealed that many of
them were actually contaminated by foreground sources
\citep{Vanzella2010a, Nestor2013, Siana2015} and indicates that a
detailed inspection of their morphologies, as well as follow-up 
spectroscopic observations are
necessary to identify genuine LyC emitting galaxies.

These previous studies have shown that the frequency of galaxies with
strong LyC emission among high-redshift star-forming galaxies (SFGs) is
relatively small; it is 5\% to $\sim$10\%
\citep[e.g.,][]{Micheva2017, Nestor2013, Steidel2018} for both
rest-frame UV colour-based LBGs and narrow-band selected Lyman $\alpha$
emitters (LAEs), although the sensitivity limits vary for different
studies. Also, the constraints on the intrinsic $f_\mathrm{esc}$ by
these studies have indicated that the average $f_\mathrm{esc}$ is
generally less than 10\%, while recent studies have found that there are 
individual SFGs with strong LyC emission
\citep{Naidu2017, Fletcher2018, Steidel2018}.
On the other hand, arguments based on the UV luminosity function of
rest-frame UV-selected SFGs at $z>6$ suggest that
$f_\mathrm{esc} \approx$10--20\% would be required for SFGs to provide a 
sufficient amount of ionizing photons for cosmic reionization
\citep[e.g.,][]{Oesch2009, Bouwens2016}.
Low $f_\mathrm{esc}$ (less than 2\%) is also suggested by independent
analyses using GRB afterglows \citep{Chen2007, Fynbo2009, Tanvir2018}.
While contributions to the ionizing photon budget from AGNs are also
necessary to be evaluated more closely
\citep{Giallongo2015, Micheva2017a, Grazian2018}, studies of LyC
emission from SFGs, especially from UV-faint galaxies, are necessary to
understand their contributions to the ionizing radiation.

Following the LyC search using Subaru / Suprime-Cam narrow-band imaging 
in the SSA22 field \citep{Iwata2009, Micheva2017},
in this paper we report the results of a search for galaxies with LyC
emission in the GOODS-N field. The GOODS-N field is one of the sky areas
which have been extensively studied, with
multi-wavelength data from X-ray to radio, and  with various deep
spectroscopic observations
\citep[e.g.,][]{Wirth2004, Barger2008, Reddy2006, Kriek2015} providing a
large number of sample galaxies at $z>3.06$ where the
Suprime-Cam \flyc filter traces rest-frame ionizing radiation.
The available deep multi-band imaging data from the \textit{Hubble}
Space Telescope (\textit{HST}) from UV to near-infrared in this field 
\citep{Giavalisco2004, Grogin2011, Koekemoer2011, Oesch2018} are
valuable to check the association of detected signal in the Suprime-Cam
narrow-band image with the $z>3$ galaxies in order to verify that the
narrow-band fluxes are LyC emission, and to discern cases with
contamination by foreground sources from LyC 
emitting galaxies. Moreover, while the SSA22 field contains a
prominent protocluster at $z=3.1$ \citep{Steidel1998, Yamada2012},
there is no known over-density of galaxies at $z>3$ in the GOODS-N. We
can examine if there is any difference in the properties related to the
LyC emission between the galaxies in a protocluster field and those in a
general field.

We structure the paper as follows.
In Section 2 we describe the narrow-band
and broad-band imaging observations,
the base sample used to search for LyC emitting galaxy candidates,
and the additional observations and data used in this study.
In Section 3, the procedure of data analyses and the selection of the
LyC emitting galaxy candidates are presented.
The properties of the LyC galaxy candidates are examined in Section 4,
and in Section 5 we discuss the implications from the results, including
a comparison with previous results in the SSA22 field and constraints on
the LyC emissivity from galaxies at $z\sim 3.1$.
Summary and conclusions are presented in Section 6.

We use a flat $\Lambda$CDM cosmology with $h_0=0.7$, $\Omega_m=0.3$.
All magnitudes are presented in the AB system \citep{OkeGunn1983}.

\section{Observations and sample construction}

\subsection{Suprime-Cam narrow-band imaging}
\label{sec:scam_n3obs}

\begin{figure}
 \includegraphics[width=\columnwidth]{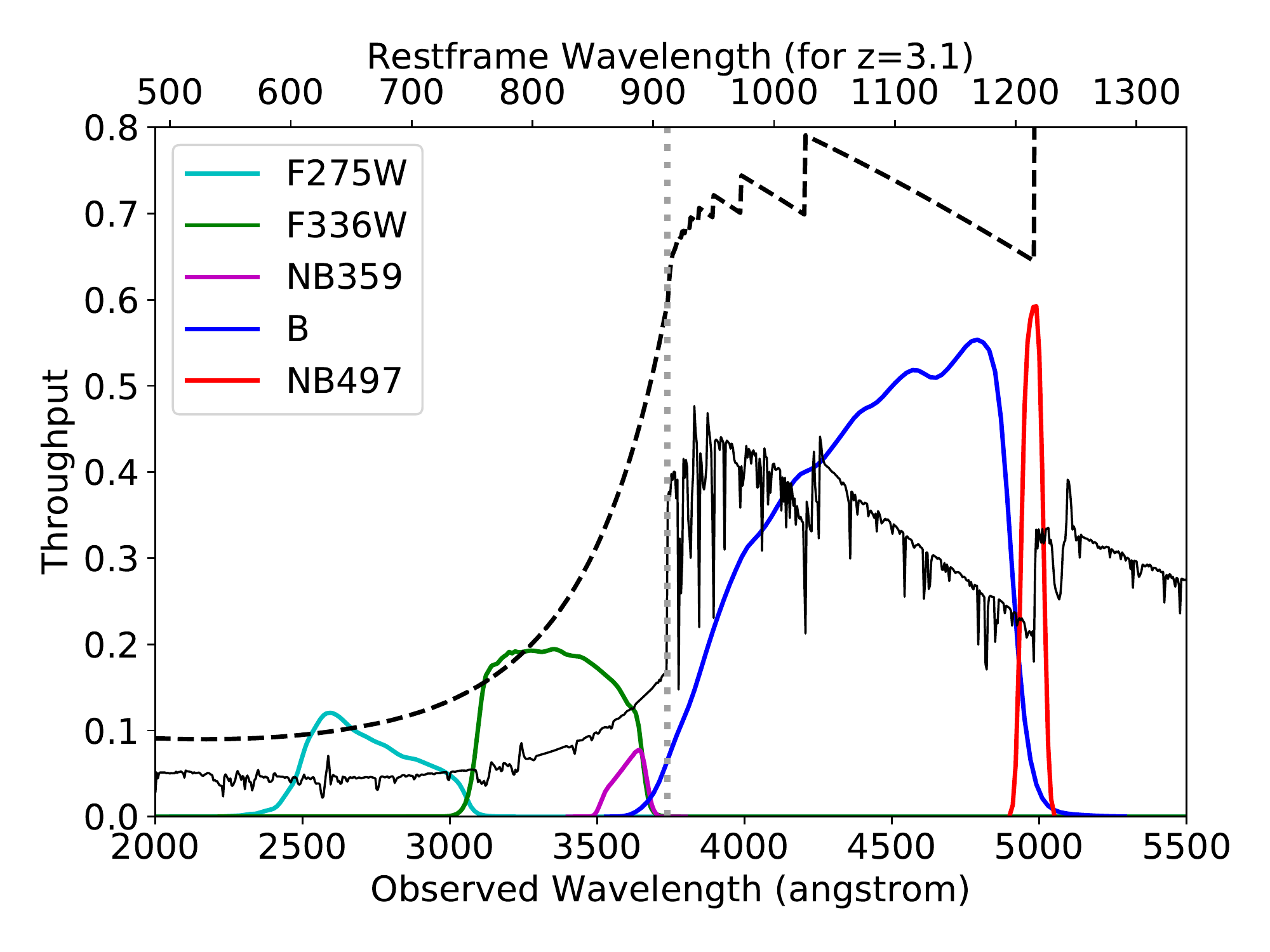}
 \caption{Transmission curves of filters for Subaru / Suprime-Cam
 and {\textit{HST}} WFC3/UVIS used in this study. Transmissions of \flyc
 (magenta), \flya (red), and B-band (blue) for Suprime-Cam include
 reflectance of the primary mirror, throughputs of Suprime-Cam and
 the atmospheric dispersion corrector of the prime-focus unit, CCD
 quantum efficiency, and typical atmospheric transmission. 
 The data for WFC3/UVIS F275W (cyan) and F336W (green) filters are taken
 from the STScI website, and also include efficiencies of other optical
 elements. 
 A model SED of a star-forming galaxy at $z=3.1$ without dust
 attenuation, with mean IGM attenuation \citep{Inoue2014}, scaled to 
 arbitrary units in $f_\lambda$, is shown with a black solid line.
 The dashed line is the mean IGM transmission at $z=3.1$, and
 the Lyman limit wavelength is indicated with a vertical dotted line.}
 \label{fig:filter_trans}
\end{figure}

We used a narrow-band filter \flyc which was designed and fabricated to 
trace LyC from galaxies at $z \gtrsim3.06$. This filter was
used in the previous study of Subaru / Suprime-Cam LyC search in the SSA22 
field \citep{Iwata2009, Micheva2017}. Laboratory measurements
have confirmed that the transmission of the filter
in the wavelength range of 400 nm -- 1200 nm is less than 0.01\%.
In Fig.\,\ref{fig:filter_trans} the transmission curves for the
Suprime-Cam filters (\textit{NB359}, \textit{NB497}, {\textit{B}}) are
shown along with the spectral energy distribution (SED) of a model
star-forming galaxy at $z=3.1$ and average IGM transmission at
this redshift. The model SED is taken from an output by BPASS version
2.1 \citep{Eldridge2017}, with a constant star formation rate, 10 Myr
age, a metallicity of $Z=0.001$, and is displayed in units of
$f_\lambda$. 
\flyc traces a narrow wavelength range just below
the Lyman limit, where relatively small IGM opacity for LyC is expected
with higher probability compared to shorter wavelengths.

The \flyc imaging observations were carried out using Suprime-Cam
\citep{Miyazaki2002}, the prime focus optical camera for the Subaru
telescope, on April 5, 2011, April 18--20, 2015, and May 7, 2016 (UT).
The field centre ($\alpha$=12:36:49.4, $\delta$=+62:08:58; J2000) was
chosen to be aligned with existing Suprime-Cam imaging observations in
broad-band \citep[e.g.,][]{Iwata2007} and \flya, and it covers the
\textit{HST}/ACS and WFC3 footprints by the GOODS \citep{Giavalisco2004}
and the CANDELS \citep{Grogin2011, Koekemoer2011}.
The total on-source integration time of \flyc imaging is 32.5 hours.

The data reduction of the \flyc images was made using SDFRED2
\citep{Ouchi2004}. Photometric calibration was made using images of
spectro-photometric standard stars taken on the same nights. After the
mosaiced image was created, astrometric registration
was made using sky coordinates of stars in the Sloan Digital
Sky Survey \citep[SDSS;][]{York2000} in the field. The rms errors of
positional alignments are $0\farcs09$--$0\farcs10$.
The median FWHM of the image for point sources after the registration is
$0\farcs95$. Five-sigma limiting magnitudes measured by randomly placing
apertures on ``empty'' sky after masking objects are 27.09 and 26.34 for
$1\farcs2$ and $2\farcs0$ diameter apertures, respectively.

\subsection{Other Suprime-Cam images}
\label{sec:scam_archive}

We use Suprime-Cam {\textit{V}}, {\textit{$I_c$}}, and
{\textit{z'}}-band imaging data of the GOODS-N field 
processed by \citet{Iwata2007}. 
In addition, we collected and processed {\textit{B}}-band
data from the archive using the same software packages as
those used for processing of the \flyc data.
The \flya image, which was used to select Ly$\alpha$ emitter
(LAE) candidates at $z=3.1$ in \citet{Yamada2012}, together
with all the Suprime-Cam images were astrometrically
registered with SDSS. 
The FWHMs of point sources and the limiting magnitudes of {\textit{B}},
{\textit{V}}, {\textit{$I_c$}}, {\textit{z'}}, \flya images are 
1\farcs02, 1\farcs11, 1\farcs13, 1\farcs13, 1\farcs15 and 
28.20, 28.17, 26.86, 26.55, 27.82 (5$\sigma$, $1\farcs2$ diameter
aperture), respectively.
For aperture photometry the PSF sizes are adjusted by convolving 
the images with better seeing sizes so that FWHMs of point sources become 
uniform over the Suprime-Cam images with different filters.

\subsection{{\textit{HST}} imaging data}
\label{sec:hst_images}

We use {\textit{HST}} archival images of the GOODS-N field from CANDELS
and HDUV \citep{Oesch2018} projects, taken from Mikulski Archive for
Space Telescopes (MAST). Although the {\textit{HST}}
observations cover a limited portion of the Suprime-Cam images as shown 
in Fig.\,\ref{fig:dist}, they have higher spatial resolution
(FWHM $\sim0\farcs09$--$0\farcs10$) than the Suprime-Cam images,
enabling us to examine the morphologies of the target galaxies in
detail.
They are also helpful to find signatures of possible foreground
contamination.
The ACS F606W and F814W images from CANDELS correspond to images in
the rest-frame non-ionizing UV wavelengths, while the WFC3/UVIS F275W
and F336W images published by HDUV, trace rest-frame wavelengths shorter
than the Lyman limit for our sample galaxies.
In Fig.\,\ref{fig:filter_trans} transmission curves of WFC3/UVIS F275W
and F336W are plotted, and the figure indicates differences in
wavelength range covered by these filters and Suprime-Cam
\textit{NB359}.

\subsection{Expected IGM transmission for ionizing photons}
\label{sec:igm_trans}

\begin{figure}
\centering
 \includegraphics[width=0.75\columnwidth]{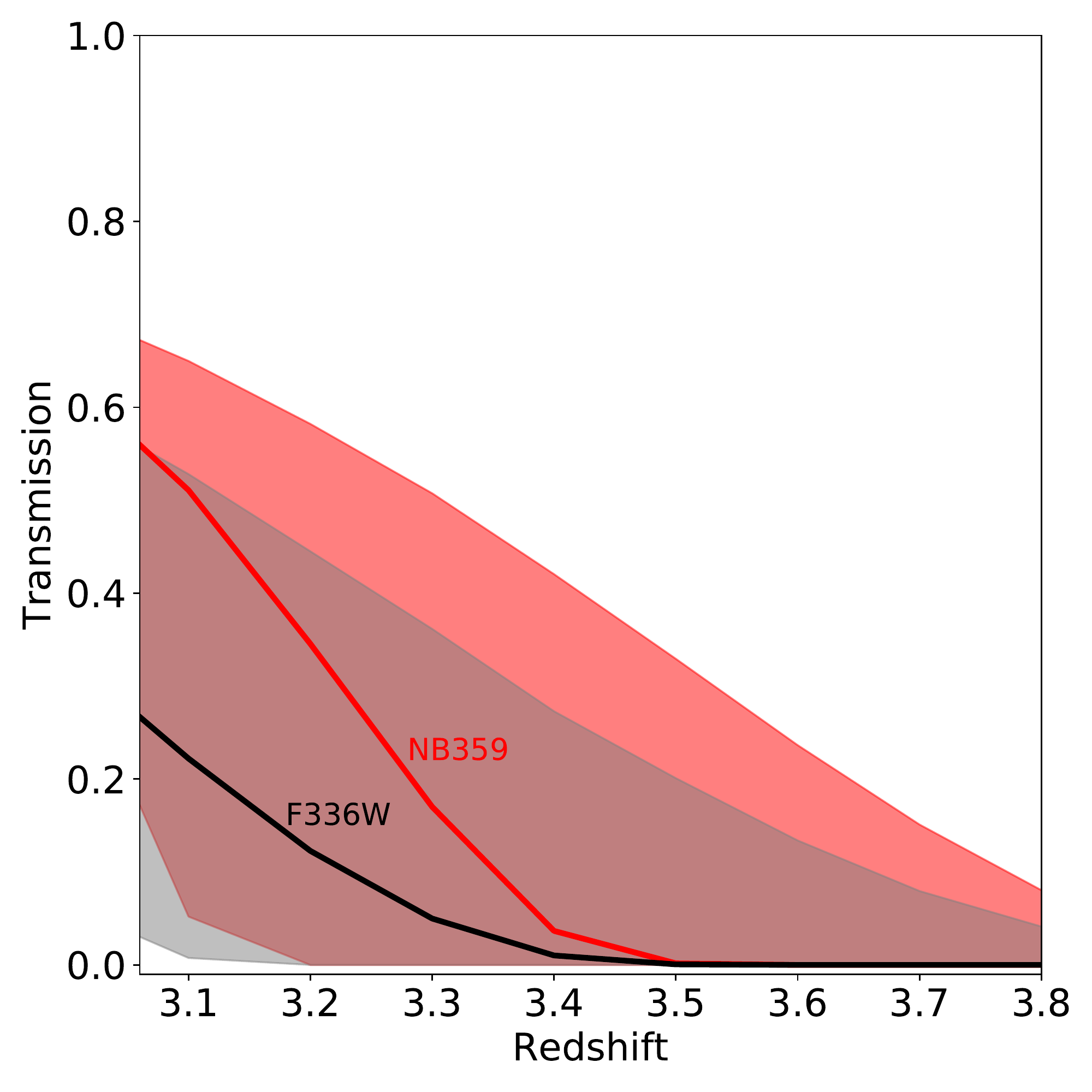}
 \caption{
 IGM transmission as a function of source redshift for Suprime-Cam
 \flyc (red) and \textit{HST} F336W (black) for a source with
 $f_{\nu}=$constant SED. Here we use a set of Monte Carlo
 simulations of 10,000 sightlines with the IGM H{\sc i} cloud
 distribution function of \citet{Inoue2014}.
 Thick lines indicate median transmission values with 0.1 redshift
 step. Filled areas represent 68\%-ile transmission values.
 }
 \label{fig:igm_trans}
\end{figure}

While both Suprime-Cam/\flyc and \textit{HST}/F336W trace LyC for
sources at $z>3.06$, \flyc has a narrower filter bandpass,
and it preferentially captures ionizing photons closer to the Lyman
limit (see Fig.\,\ref{fig:filter_trans}). Because IGM attenuation varies
as a function of wavelength, expected transmissions of photons through
the IGM (here we denote the ratio of flux density with and without
IGM attenuation as `IGM transmission') are different for different
filters. In Fig.\,\ref{fig:igm_trans} we show the median 
and 68\%-ile IGM transmission values of \flyc and F336W for
a source at a redshift range of $3.1\leq z \leq 3.8$.
These IGM transmissions are estimated from Monte Carlo simulations
\citep{Inoue2008} which generate 10,000 sightlines for redshifts
consistent with the H{\sc i} cloud distribution defined analytically by
\citet{Inoue2014}.
For each sightline we calculate IGM transmission through \flyc or F336W
for an object with a flat SED in $f_\nu$ (i.e., $f_\nu$=constant)
and obtain median and 68\%-ile values from the 10,000
instances. These calculations are repeated from $z=3.1$ to 3.8 with a
redshift step of 0.1.
The median values of IGM transmission for \flyc are higher than those
for F336W, especially at the lower redshift range. This is because 
\flyc traces a rest-frame wavelength range close to the Lyman limit
where IGM transmission is higher than that for photons with shorter
wavelengths. 
If the same limiting magnitude is achieved, \flyc is more sensitive for
detecting LyC photons from $z>3$ galaxies than F336W.
Median values of IGM transmission for both \flyc and F336W rapidly
decrease as the source redshift increases, which indicates that the
detection of LyC photons becomes increasingly difficult due to
increasing IGM opacity. However, higher-side 68\%-ile IGM transmission
values at $z=3.5$ are 0.33 for \flyc and 0.20 for F336W, respectively. 
The fluctuation of IGM transmission among different sightlines is large,
and there is still a reasonable number of sightlines with significant
IGM transparency for LyC in the redshift range studied in this paper. 

\citet{Rudie2013} used a correlation of the Ly$\alpha$ forest in spectra
of high-$z$ QSOs and galaxies in projected surrounding areas of the QSOs
and found higher frequency of H{\sc i} clouds with column densities
$14\lesssim$ log($N_{\mathrm{HI}}$/cm$^{-2})$ $\lesssim 17.2$ in
the CGM ($\leq$300 pkpc) of $2<z<3$ galaxies compared to the IGM.
\citet{Rudie2013} and \citet{Steidel2018} showed that the average LyC
transmission is reduced when attenuation by the CGM in addition to the
IGM is considered. Thus, the expected transmission of LyC shown in
Fig.\,\ref{fig:igm_trans} could be overestimated if the galaxies in
question are accompanied by the CGM, which contains H{\sc i} clouds with
higher number density than the IGM. We will discuss such an effect of
the CGM when we examine the escape fraction and emissivity of LyC for
our sample galaxies in Section~\ref{subsec:lyc_emissivity}.

\subsection{Sample selection}
\label{sec:sample_selection}

The $z=3.1$ LAE sample comes from the Subaru / Suprime-Cam \flya 
imaging observations \citep{Yamada2012}.
The selection was made based on the excess of \flya flux densities over
those in broad-band filters, which corresponds to the observed
Ly$\alpha$ equivalent widths higher than $\approx$190\AA\, 
(see \citet{Yamada2012} for details).
The number of LAE candidates in the field with the \flyc image is 158.
These LAE candidates in the GOODS-N field have not been
spectroscopically followed-up, other than a small subsample observed
with Subaru / MOIRCS in the present study (described in Section
\ref{sec:moircs_obs}). 
There was only one object with a spectroscopically confirmed redshift, a
known quasar SDSS J123557.63+621024.4 at $z=3.075$ \citep{Wirth2004}
which is included in our `AGN' sample and is removed from the `LAE'
sample.

We compiled a sample of galaxies with spectroscopic redshifts from
literature. The lower limit of the redshift range was set to be 3.06,
above which the \flyc filter transmission is free from contamination 
by non-ionizing photons. We set the upper limit of our primary sample to
be $z=3.5$, at which redshift the central wavelength of the \flyc
filter corresponds to $\simeq800$\AA,
and flux from an object at a redshift higher than that through \flyc
will frequently suffer heavy IGM attenuation (see
Fig.~\ref{fig:igm_trans}).
Most of our sample galaxies come from \citet{Reddy2006} which provides a
catalog of Lyman break galaxies (LBGs) in the GOODS-N field, and MOSDEF
\citep[MOSFIRE Deep Evolution Field;][]{Kriek2015}.
We also considered \citet{Wirth2004} and \citet{Barger2008} as
compilations of redshifts in the GOODS-N, as well as more recent reports
of spectroscopic observations in the field. 
In Table~\ref{tab:sample_src} we summarize the sources of the sample
galaxies with spectroscopic redshift between 3.06 and 3.5.
Because the spectroscopic redshift information comes from different
observations, the quality and precision of the redshifts
may not be uniform. We excluded several objects from the literature for
which the authors marked their redshifts as uncertain.
For each galaxy in the list we identified a counterpart in the GOODS
version 2 ACS catalog and checked the photometric redshifts and HST WFC3 
grism data from 3D-HST \citep{Skelton2014,Momcheva2016}. 
Occasionally, a peak photometric redshift in the 3D-HST photometric
catalog is lower than our redshift lower limit of 3.06. Even in such
cases we adopt the spectroscopic redshift in the literature if no
quality warning is given in the catalog. In the WFC3 grism data, if 
emission lines are significantly detected, the estimated redshift is
mostly consistent with the ground-based spectroscopic redshift. However, 
there is one case for which the grism redshift ($z=2.01$) and the
spectroscopic redshift ($z=3.226$) are largely different while the
spectroscopic redshift is marked as robust in the spectroscopic catalog
\citep{Kriek2015}. This object is separated from the main sample and it
will be discussed in Section~\ref{subsec:nbphot}.
In total there are 103 galaxies with known spectroscopic redshifts in
the range $3.06<z<3.5$, and all of them are within the ACS coverage of
the GOODS-N field.

We also investigate LyC from AGNs in the same redshift range. In
addition to four objects in the Subaru / Suprime-Cam field reported as
AGNs in the literature 
\citep{Hornschemeier2001, Barger2002, Reddy2006, Shen2007},
we cross-checked galaxies with spectroscopic redshifts with the Chandra
X-ray catalogs \citep{Alexander2003, Xue2016} and included four galaxies
with an X-ray detection. All of them are categorized as AGN in the
catalog by \citet{Xue2016}. Although it is not evident whether UV
radiation from these relatively less luminous objects originates from
AGN or star-formation, in this study these objects are included in the
`AGN' sample and are excluded from the sample of SFGs.
The total number of objects in the AGN sample is eight. 
Two among them are located outside of the GOODS-N ACS coverage.

\begin{table}
 \centering
 \caption{Numbers and origins of the sample star-forming galaxies with
 known spectroscopic redshift $3.06 < z < 3.5$ in the GOODS-N field.}
 \label{tab:sample_src}
 \begin{tabular}{lr}
  \hline
  Source & Number \\
  \hline
  \citet{Kriek2015}     & 41\\
  \citet{Reddy2006}     & 34\\
  \citet{Schenker2013}  & 9 \\
  \citet{Momcheva2016}  & 7\\
  \citet{Lowenthal1997} & 3\\
  \citet{Adams2011}     & 3\\
  \citet{U2015}         & 2\\
  \citet{Cohen2000}     & 1\\
  \citet{Shapley2001}   & 1\\
  \citet{Wirth2004}     & 1\\
  \citet{Pirzkal2013}   & 1\\
  \hline
  Total & 103 \\
  \hline
 \end{tabular}
\end{table}

\begin{figure*}
\centering
 \includegraphics[width=0.80\columnwidth]{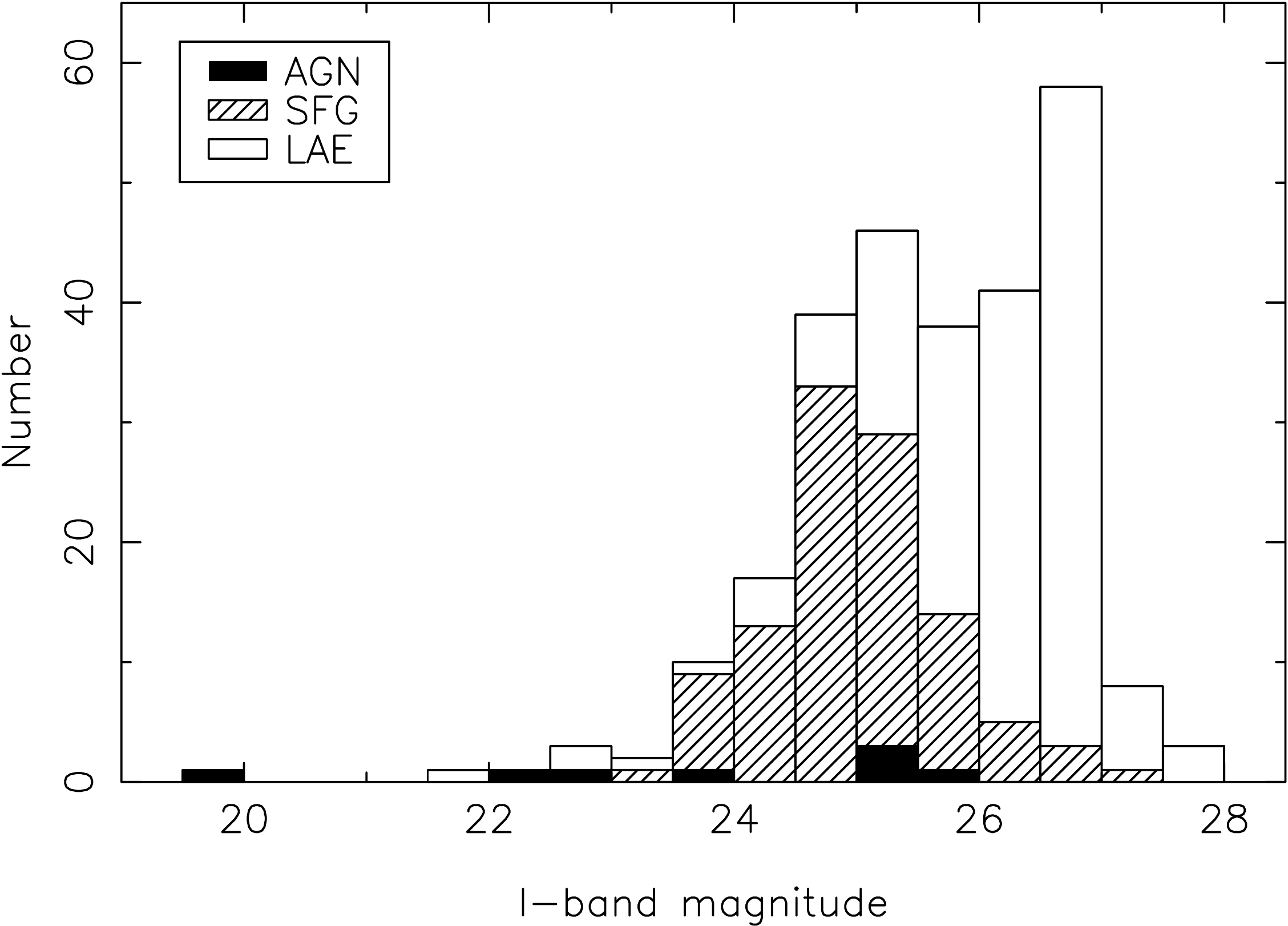}
 \hspace{0.1\columnwidth}
 \includegraphics[width=0.80\columnwidth]{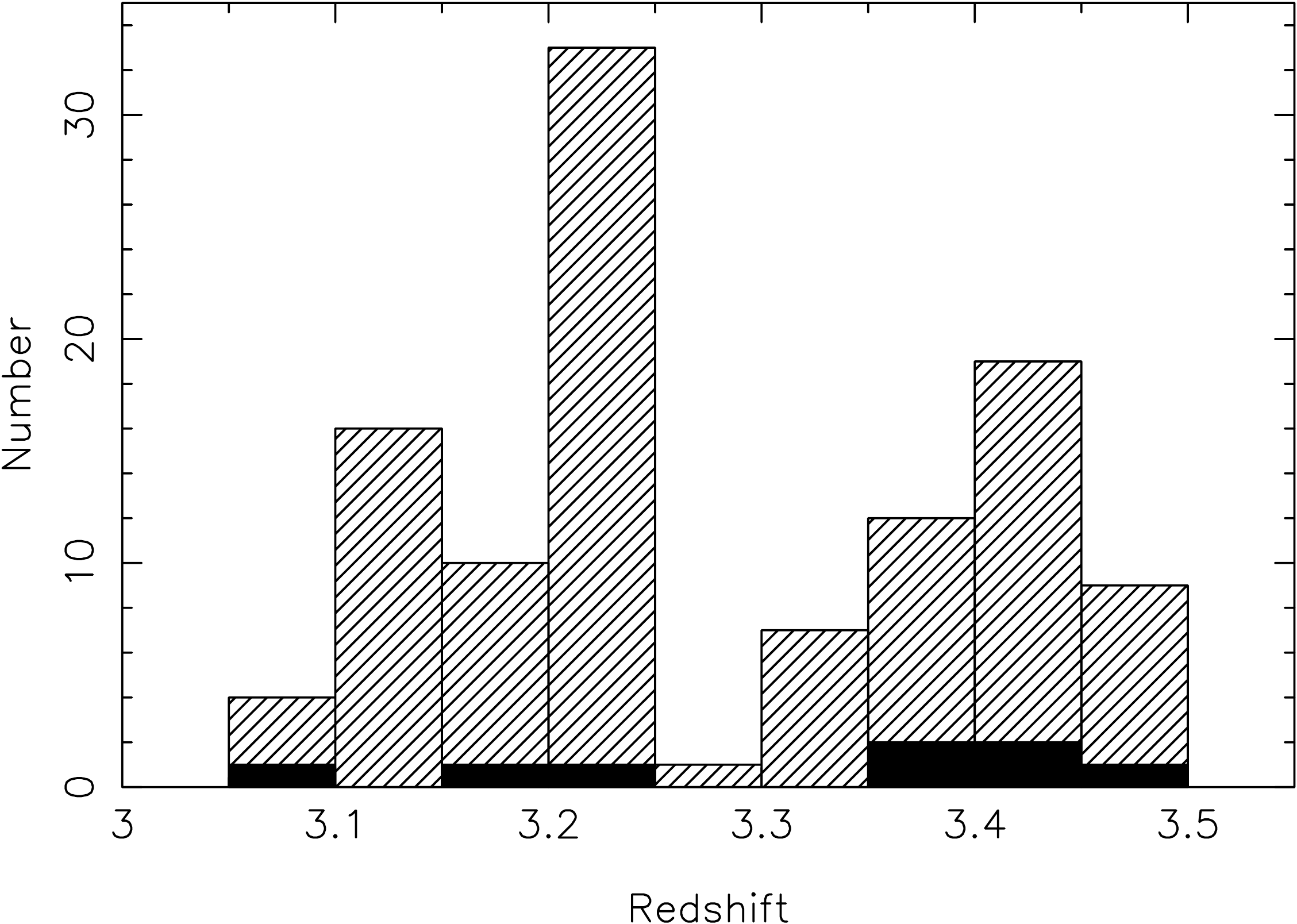}
 \caption{{\textit{Left}}: $I_c$-band magnitude distribution of the
 sample galaxies. Filled, hatched, and open histograms show numbers of
 AGNs, SFGs, and LAEs, respectively.
 {\textit{Right}}: redshift distribution of AGN (filled) and SFG
 (hatched) sample galaxies.
}\protect \label{fig:imaghist}
\end{figure*}

\begin{figure*}
 \includegraphics[width=150mm]{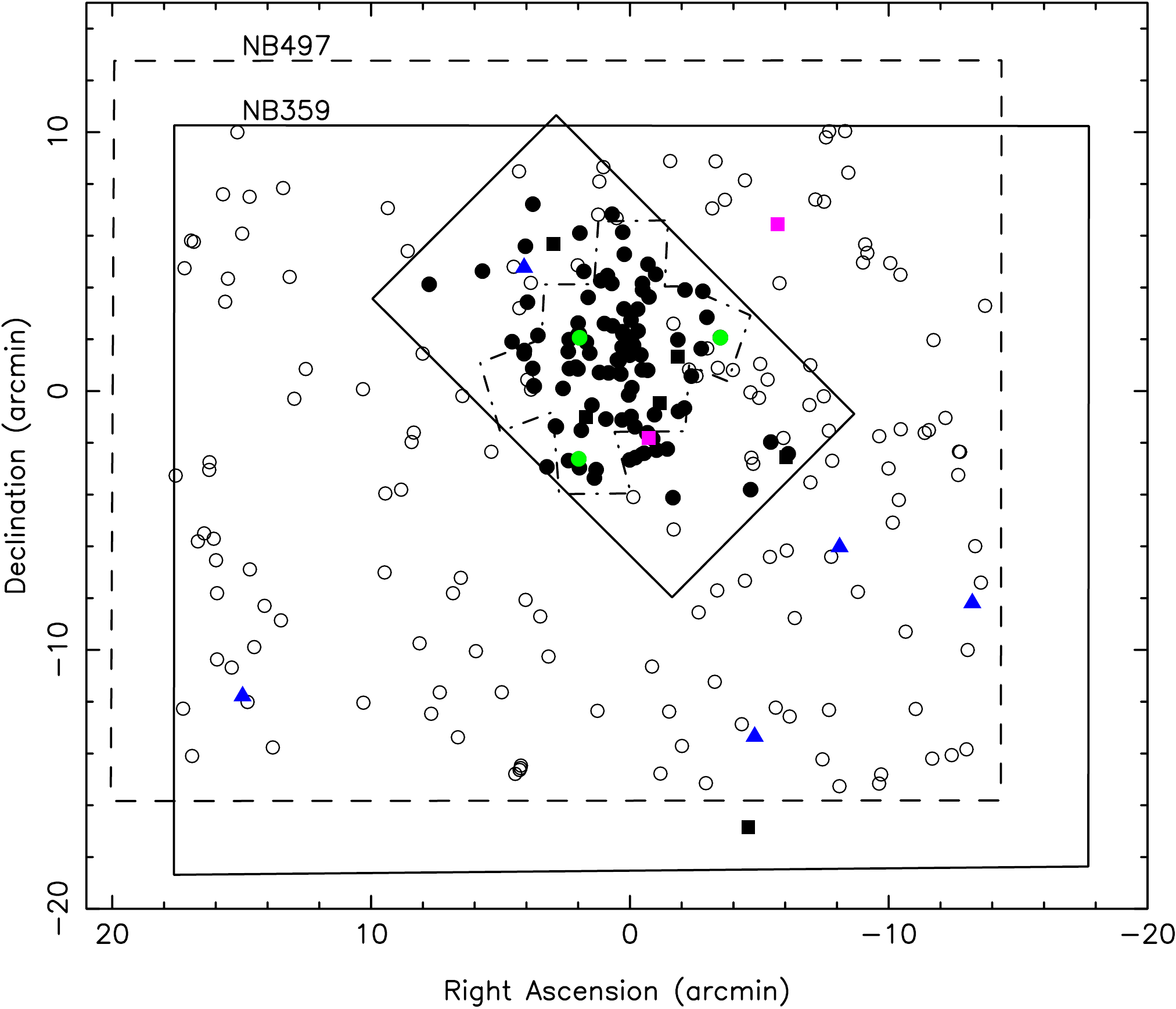}
 \caption{Spatial distribution of the sample galaxies. The origin of the
 coordinates is the position of the Hubble Deep Field -- North
 ($\alpha$=12:36:49.4, $\delta$=+62:12:58; J2000). The inner solid
 rectangle represents {\textit{HST}} ACS coverage, and the dot-dashed line is
 the area covered by HDUV \citep{Oesch2018} with {\textit{HST}} WFC3/UVIS
 F336W. The larger solid and dashed rectangles show the coverage of
 Subaru Suprime-Cam with the \flyc and \flya\ filters,
 respectively. Small filled squares, filled circles, and open circles 
 represent the positions of AGNs, SFGs, and LAEs in the sample,
 respectively.
 Coloured markers show positions of the LyC detection candidates
 (green circles: SFGs, magenta squares: AGNs, blue triangles: LAEs).}
 \label{fig:dist}
\end{figure*}

The $I_c$-band magnitude distribution of the sample galaxies (including
AGNs and LAEs) and the redshift distribution of the SFGs and AGNs are
shown in Fig.\,\ref{fig:imaghist}.
In Fig.\,\ref{fig:dist} we show a spatial distribution of the sample as
well as the field coverage of the {\textit{HST}} and Suprime-Cam images.

\subsection{MOIRCS spectroscopy}
\label{sec:moircs_obs}

We conducted follow-up near-infrared multi-object spectroscopy
observations for some of the LyC emitting galaxy candidates using the 
Multi-Object Infrared Camera and Spectrograph
\citep[MOIRCS;][]{Suzuki2008} at the Subaru Telescope which has recently
received a detector upgrade with two Teledyne H2RGs
\citep{Walawender2016, Fabricius2016}.
The observations were on March 2, May 21, and June 17, 2018 (UT) with
clear sky and moderately good seeing condition. In total, three fields in
the GOODS-N area have been observed. Details of the observations are
given in Table~\ref{tab:moircs_obs}.
For all these observations we used the HK500 grism which covers 1.3--2.3
$\mu{\mathrm m}$ in a single exposure and delivers a spectral resolution
of $R\approx300$ with 0\farcs9 slits.

The data reduction was made using NOAO/IRAF, with some commands from
MCSMDP\footnote{MCSMDP is distributed in the MOIRCS web site at
\url{https://www.subarutelescope.org/Observing/Instruments/MOIRCS/}},
the multi-object spectroscopy data reduction package for MOIRCS
\citep{Yoshikawa2010}.

\begin{table*}
 \centering
 \caption{Details of MOIRCS observations.}
 \label{tab:moircs_obs}
 \begin{tabular}{lcrc}
  \hline
  Pointing centre (J2000) & Slit width($\arcsec$) & Exposure Time & Observing date (UT) \\
  \hline
  12:36:53.179 +62:12:44.25 & 0.9 & 10440 sec & March 2, 2018 \\
                            & 0.7 &  8640 sec & May 21, 2018\\
  12:35:35.960 +62:09:26.04 & 0.9 &  7200 sec & March 2, 2018\\
  12:36:41.080 +62:14:56.10 & 0.7 &  4680 sec & June 17, 2018\\
  \hline
 \end{tabular}
\end{table*}

\section{Selection of LyC candidates}
\label{sec:lyc_selection}

\subsection{Narrow-band image photometry and selection of LyC candidates}
\label{subsec:nbphot}

We applied aperture photometry to the \flyc image at the centroid
positions of the sample objects obtained from the {\textit{z}}-band
(F850LP for ACS) for the SFGs and from \flya\ for the LAEs.
For SFGs within the ACS GOODS-N coverage, the positions were taken
from the v2.0 of the GOODS ACS catalogue
\footnote{For one object in the SFG
sample (MOSDEF 19947) the GOODS catalog does not separate it from a
nearby bright source and the object is not recorded in the catalog. For
this object we applied the aperture photometry at its peak position in
F850LP image.}.
We used a $1\farcs2$ diameter aperture to find candidates of LyC
emission.
We should note that this procedure is different from the way LyC emitter
candidates are identified in \cite{Iwata2009} and
\cite{Micheva2017}, where sources with $\geq$3$\sigma$ detection in
\flyc with a $1\farcs2$ diameter aperture within a $1\farcs2$ (in
\cite{Iwata2009}) or $1\farcs4$ (in \cite{Micheva2017}) radius of the 
{\textit{R}}-band position are selected. As discussed in
\cite{Micheva2017}, the procedure in the previous studies allows a
detection of LyC emission with a spatial offset from the rest-frame 
non-ionizing UV radiation, which may happen in highly active,
morphologically disturbed galaxies or with star formation taking place
in the outskirts of the galaxies.
However, because half-light radii or effective radii of $z\sim 3$
star-forming galaxies are typically less than 2 kpc 
\citep[e.g.,][]{Bouwens2004, Shibuya2015}, $1\farcs2$ or
$1\farcs4$ (which corresponds to $\approx 9$ and 11 kpc, respectively)
may be too large to be a threshold of physical association. Moreover,
it also increases the chances of mis-identification of sources
with overlapping foreground objects, which have been recognized as a
major obstacle in finding genuine LyC emitters 
\citep[e.g.,][]{Vanzella2010a, Siana2015}. In the present study, in
order to reduce the probability of chance overlap with foreground
sources, we choose galaxies with $\geq$3$\sigma$ signal in \flyc at the
centroid position in the rest-frame UV wavelength.

\begin{figure}
 \includegraphics[width=\columnwidth]{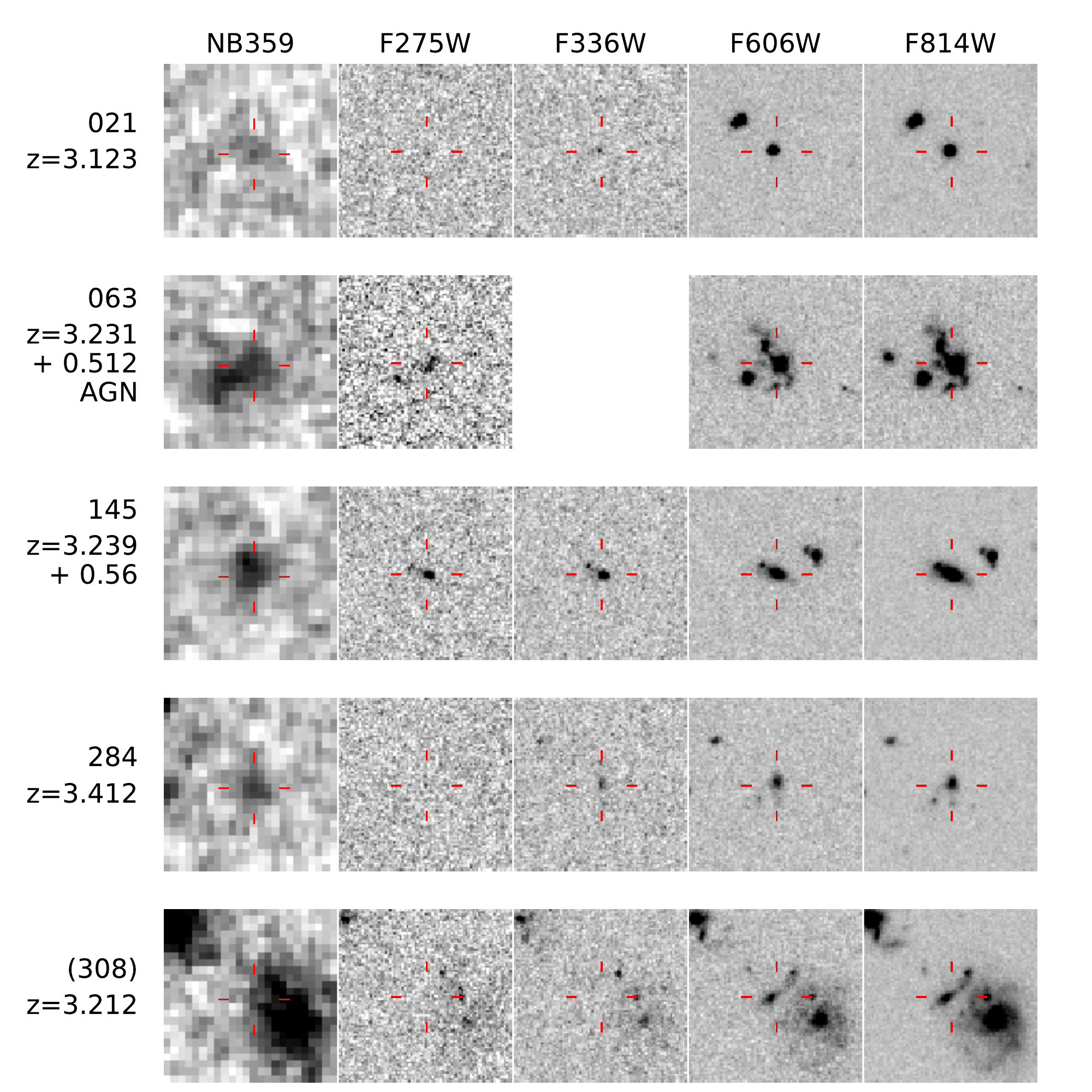}
 \caption{Subaru / Suprime-Cam \flyc and {\textit{HST}} WFC3/UVIS and
 ACS images of \flyc detected sources with spectroscopic redshift
 between 3.06 and 3.5. Each postage stamp image has 
 $5\arcsec \times 5\arcsec$ size.
 Objects 063 and 145 have spectroscopic interlopers reported by Jones et
 al.~(2018) and their redshifts are also listed as $+0.512$ and
 $+0.560$, respectively.
 Object 063 lies outside of F336W coverage of HDUV.
 The \flyc photometry of object 308 is affected by a nearby source, and
 this object is excluded from the LyC emitting galaxy candidates.}
 \label{fig:montage_SFGAGN}
\end{figure}

In the SFG sample, four among 103 sample galaxies have
$\geq$3$\sigma$ signal at the centroid position in the {\textit{z}}-band
image. In Fig.\,\ref{fig:montage_SFGAGN} we show the postage stamp
images of the four SFGs in Subaru / Suprime-Cam \textit{NB359},
{\textit{HST}} WFC3/UVIS F275W and F336W from HDUV, and {\textit{HST}}
ACS F606W and F814W from CANDELS. For object 308 (MOSDEF 19947) the
\flyc image is contaminated by a nearby bright galaxy. We treat this as
a false detection and the object is removed from the LyC candidates. 

\begin{figure}
 \includegraphics[width=\columnwidth]{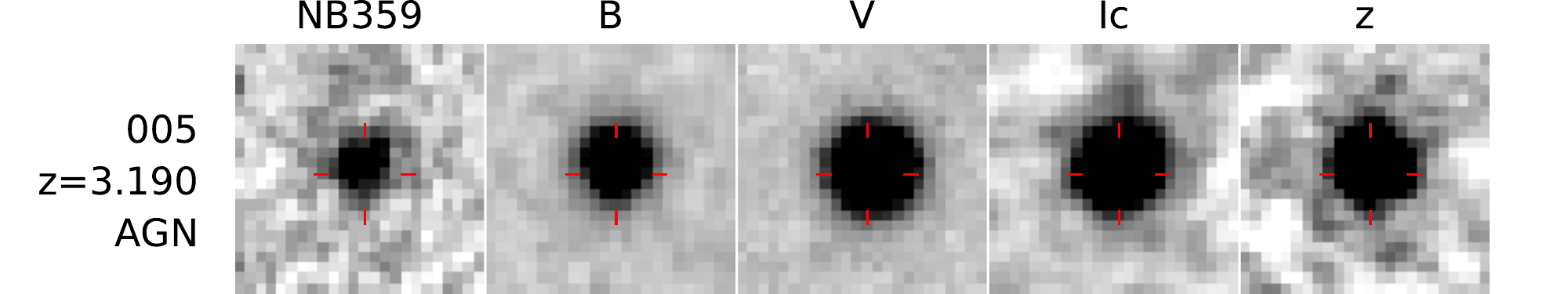}
 \caption{Subaru / Suprime-Cam $5\arcsec \times 5\arcsec$ postage stamp
 image of Object 005 (B02-049), an AGN at $z=3.19$ with
 \flyc detection.}
 \label{fig:montage_AGN005}
\end{figure}

Among the eight objects in the AGN sample, six are within the GOODS-N
field. One object with running number 063 has $\geq$3$\sigma$ signal.
The postage stamp image of the object is shown in
Fig.\,\ref{fig:montage_SFGAGN}. Another object with $\geq$3$\sigma$
\flyc signal among the AGN sample is B02-049, shown in
Fig.\,\ref{fig:montage_AGN005}. This object is outside of the GOODS-N
ACS coverage. 
Table~\ref{tab:lyc_properties} summarizes some basic information on the
SFG/AGN objects with \flyc signal detection.
Among these five objects, \cite{Jones2018} obtained Keck / DEIMOS
optical spectroscopy of object 145: R06-BX1400 (their GN-UVC-3) and
object 063: MOSDEF~08780 (their GN-UVC-2), and found emission lines
indicating contamination by lower redshift sources at $z=0.560$ and
$z=0.512$, respectively.
We will further discuss the possibility of foreground contamination of
the SFG LyC candidates in Section~\ref{subsec:lyc_reality}.

\begin{figure}
 \includegraphics[width=\columnwidth]{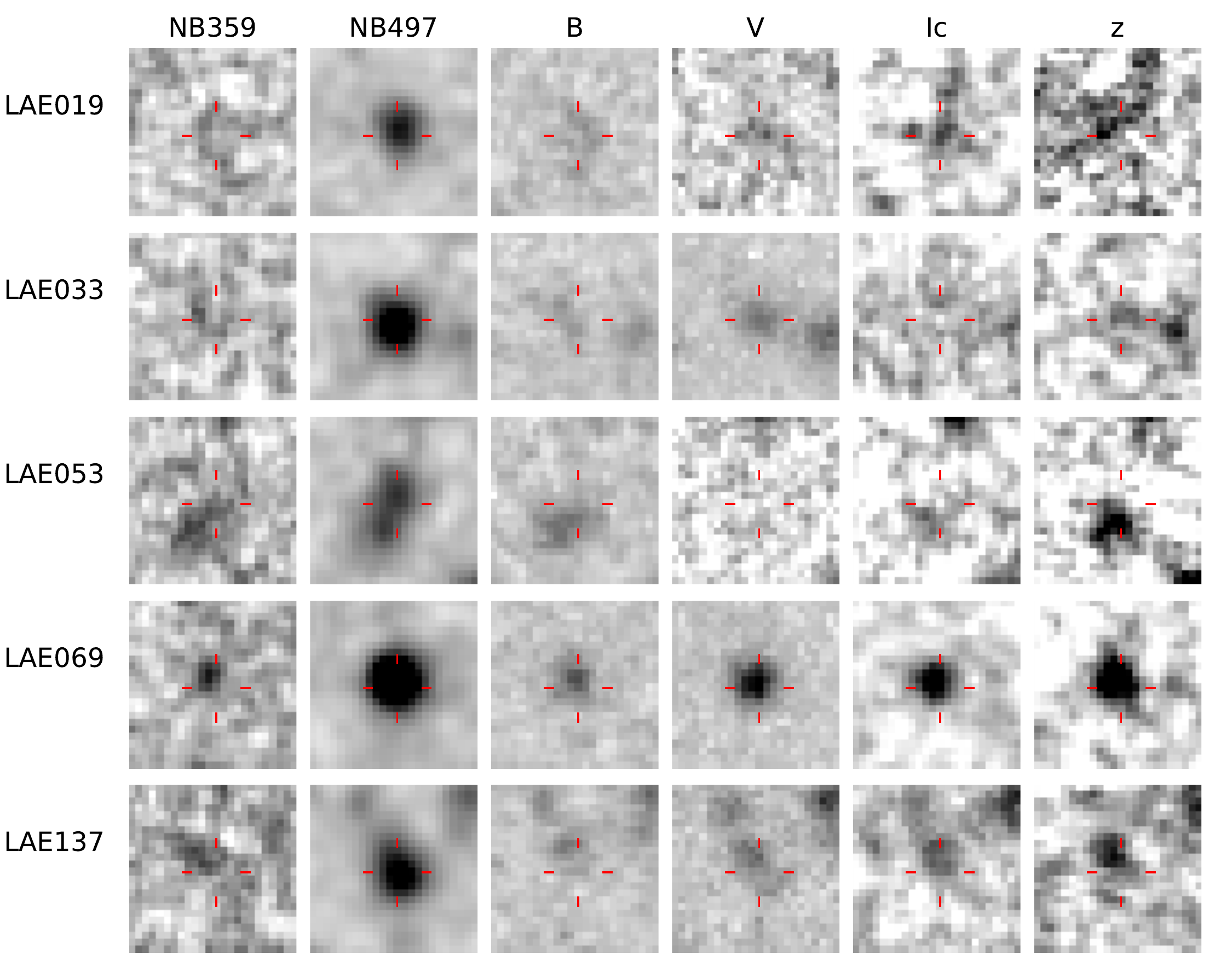}
 \caption{Subaru / Suprime-Cam $5\arcsec \times 5\arcsec$ postage stamp
 images of \textit{NB497}-selected LAE candidates with \flyc
 detection.}
 \label{fig:montage_LAE}
\end{figure}

\begin{figure}
 \includegraphics[width=\columnwidth]{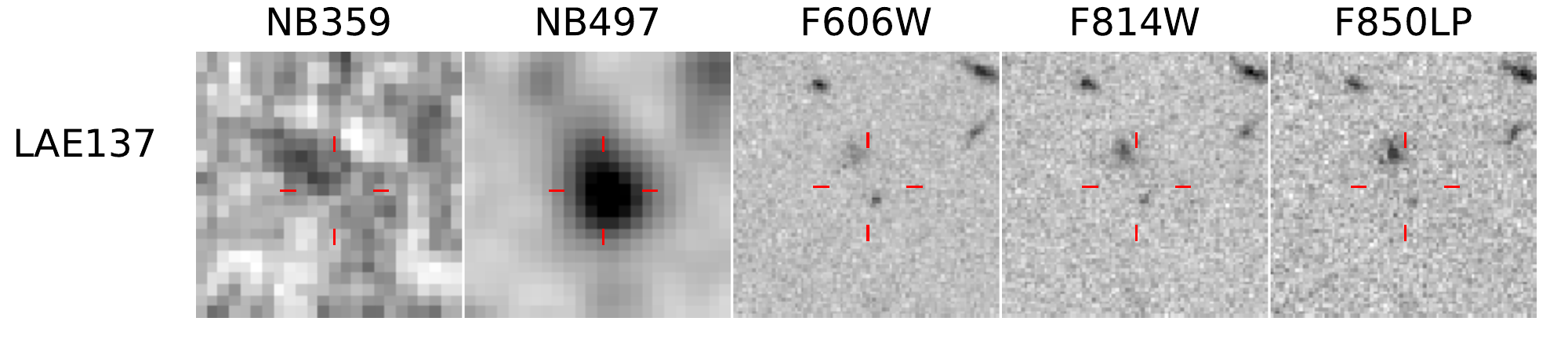}
 \caption{Subaru / Suprime-Cam \textit{NB359}, \flya and
{\textit{HST}}  ACS images of LAE~137, the only LAE candidate within the
 GOODS-N {\textit{HST}} coverage with \flyc detection.
 The red hairline indicates the
 position of Suprime-Cam \flya image centroid.}
 \label{fig:montage_LAE137_HST}
\end{figure}

Among 157 objects in the LAE sample (with the one known quasar
eliminated from the sample), five objects have $\geq$3$\sigma$ signal
in the \flyc filter at the peak positions in the \flya\ filter with a
$1\farcs2$ diameter aperture. Postage stamp images of these objects are
shown in Fig.~\ref{fig:montage_LAE}, and Table~\ref{tab:lyc_properties}
provides their basic information.
As shown in the spatial distribution (Fig.~\ref{fig:dist}), our LAE
sample distributes over the entire Suprime-Cam image, and among these
six objects only one object, LAE~137 is located within the
{\textit{HST}} / ACS coverage. The ACS image of the object
(Fig.~\ref{fig:montage_LAE137_HST}) shows that there are in fact two  
objects, one associated with the \flyc and the other with the \flya
signals. Therefore, the high spatial resolution ACS image suggests that
this \flyc detection may be due to foreground contamination. Similarly,
for LAE~053 in the seeing-limited Suprime-Cam images in
Fig.~\ref{fig:montage_LAE} there are two knots in its \flya image, and
the \flyc emission is associated only with one of the \flya knots
which is seen in broad-band images but is not associated with the other
knot only visible in \textit{NB497}. This geometry suggests that this is
another instance of foreground contamination.
The spatial offsets between the centroid in \flyc and that in \flya for
these two objects are $>0\farcs5$ which are larger than the offset
values for the other candidates (Table~\ref{tab:lyc_properties}).
We also treat another object, LAE 033 as another LAE 
possibly being contaminated by a foreground object, based on the large 
\flycfuv value (1.30). We will make more discussion in
Section~\ref{sec:reality_LAE} on possibilities of foreground
contamination.

\begin{table*}
 \centering
\caption{List of \flyc detected sources among objects with
 spectroscopic redshifts between 3.06 and 3.5 and $z=3.1$ LAE candidates.
}
\label{tab:lyc_properties}
\begin{tabular}{clcclcccll}
\hline
    & Designation  & R.A. & Decl. & Redshift & $m_{\mathrm{NB359}}^a$ & $m_{I_c}^a$ &  $f_{\mathrm{LyC}}/f_{\mathrm{UV}}^b$ & Sep.$^c$ & Ref.$^d$ \\
 \cline{3-4}
    & & \multicolumn{2}{c}{(J2000)} & & & & & ($''$) & \\
\hline
\multicolumn{10}{c}{Star-forming galaxies}\\
\hline
021 & R06-D23        & 189.08048 & 62.25039 & 3.123 & $27.08\pm0.27$ & $24.411\pm0.026$ & $0.12\pm0.03$ & $0.08$ & 1 \\
145 & R06-BX1400$^e$ & 189.27555 & 62.25037 & 3.239 + 0.560 & $26.02\pm0.16$ & $24.335\pm0.024$ & $0.21\pm0.03$ & $0.06$ & 1 \\
284 & MOSDEF 06336   & 189.27660 & 62.17246 & 3.412 & $26.48\pm0.20$ & $26.243\pm0.103$ & $0.97\pm0.27$ & $0.06$ & 2 \\
\hline
\multicolumn{10}{c}{AGNs}\\
\hline
005 & B02-049          & 189.00165 & 62.32369 & 3.190  & $25.46\pm0.12$ & $24.078\pm0.021$ & $0.25\pm0.03$ & $0.18$ & 3 \\
063 & MOSDEF 08780$^f$ & 189.17954 & 62.18572 & 3.2306 + 0.512 & $25.95\pm0.15$ & $23.599\pm0.017$ & $0.61\pm0.20$ & $0.41$ & 2 \\
\hline
\multicolumn{10}{c}{LAEs}\\
\hline
& LAE~019     & 189.03346 & 61.99364 & -- & $27.07\pm0.26$ & $26.50\pm0.14$ & $0.59\pm0.16$ & $0.20$ &\\
& LAE~033$^g$ & 189.74088 & 62.01967 & -- & $27.11\pm0.27$ & $27.39\pm0.25$ & $1.30\pm0.42$ & $0.31$ &\\
& LAE~053$^g$ & 188.73271 & 62.07962 & -- & $26.70\pm0.22$ & $27.40\pm0.37$ & $1.89\pm0.71$ & $0.81$ &\\
& LAE~069     & 188.91621 & 62.11565 & -- & $26.73\pm0.23$ & $25.75\pm0.06$ & $0.41\pm0.09$ & $0.25$ &\\
& LAE~137$^g$ & 189.35200 & 62.29561 & -- & $26.88\pm0.24$ & $26.89\pm0.15$ & $1.01\pm0.26$ & $0.59$ &\\
\hline
\end{tabular}
\begin{flushleft}
\textit{Notes.}
$^a$: Magnitudes are measured with a $1\farcs6$ diameter aperture.
$^b$: $f_{\mathrm{LyC}}$ is the flux density in \flyc, and
 $f_{\mathrm{UV}}$ is the flux density at rest-frame 1500\AA\
 (see text for details.)
$^c$: Spatial offsets between the centroid in \flyc and that in the
 $I_c$-band (for SFGs and AGNs) or in \flya (for LAEs).
$^d$: Redshift references: (1) \citet{Reddy2006}, (2) \citet{Kriek2015},
(3) \citet{Barger2002}.
$^e$: GN-UVC-3 with a spectroscopic interloper at $z=0.560$ in \citet{Jones2018}.
$^f$: GN-UVC-2 with a spectroscopic interloper at $z=0.512$ in \citet{Jones2018}.
$^g$: possibly contaminated by a foreground source (see text).
\end{flushleft}
\end{table*}

In Table~\ref{tab:lyc_properties} we list the AB magnitudes in \flyc and
$I_c$  of the \flyc detected sources. Although during the selection of
\flyc detected objects we use a $1\farcs2$ diameter aperture, for the
photometry data shown in the table we use a $1\farcs6$ diameter
aperture, which gives on average about 40\% larger flux densities with
Suprime-Cam broad-band filters than the $1\farcs2$ diameter aperture, so
that these flux densities and flux ratios represent values closer to the
`total' values. 
We also show the flux ratios between LyC and non-ionizing UV photons at
rest-frame 1500~\AA\ (labeled as `\flycfuv').
The \flyc flux density is used for $f_{\mathrm{LyC}}$.
For SFGs and AGNs, we calculate rest-frame 1500~\AA\ flux densities from
the flux densities in {\textit{$I_c$}}-band and {\textit{z}}-band, with
a simple linear extrapolation:
\begin{equation}
 f_{\mathrm{UV}} = (f_{I_c}-f_{z}) \frac{1500 (1+z) - \lambda_{I_c}}{\lambda_{I_c}-\lambda_{z}} + f_{I_c},
\end{equation}
where $f_{I_c}$ and $f_z$ are the flux densities in $I_c$-band and
$z$-band, respectively, $z$ is the redshift of the object, and
$\lambda_{I_c}$ and $\lambda_z$ are the effective wavelengths of the
filters (7979\AA\ and 9067\AA, respectively). Here we do not use the
flux density in the $V$-band to estimate the rest-frame UV slope,
because Ly$\alpha$ photons could be contained within the $V$-band filter
for an object in the redshift range considered in the present study.
Since the {\textit{$I_c$}}/{\textit{z}} flux ratios of the SFGs and AGNs
are close to unity, the changes in flux ratios when simply using $I_c$
flux densities are small and they are within the errors given in the 
table.
Since the continuum flux densities of LAEs are low, and their rest-frame
UV slopes are not well constrained, we use their flux densities in
{\textit{$I_c$}} which correspond to $\approx1950$~\AA\ in rest-frame. 

\begin{figure}
 \includegraphics[width=\columnwidth]{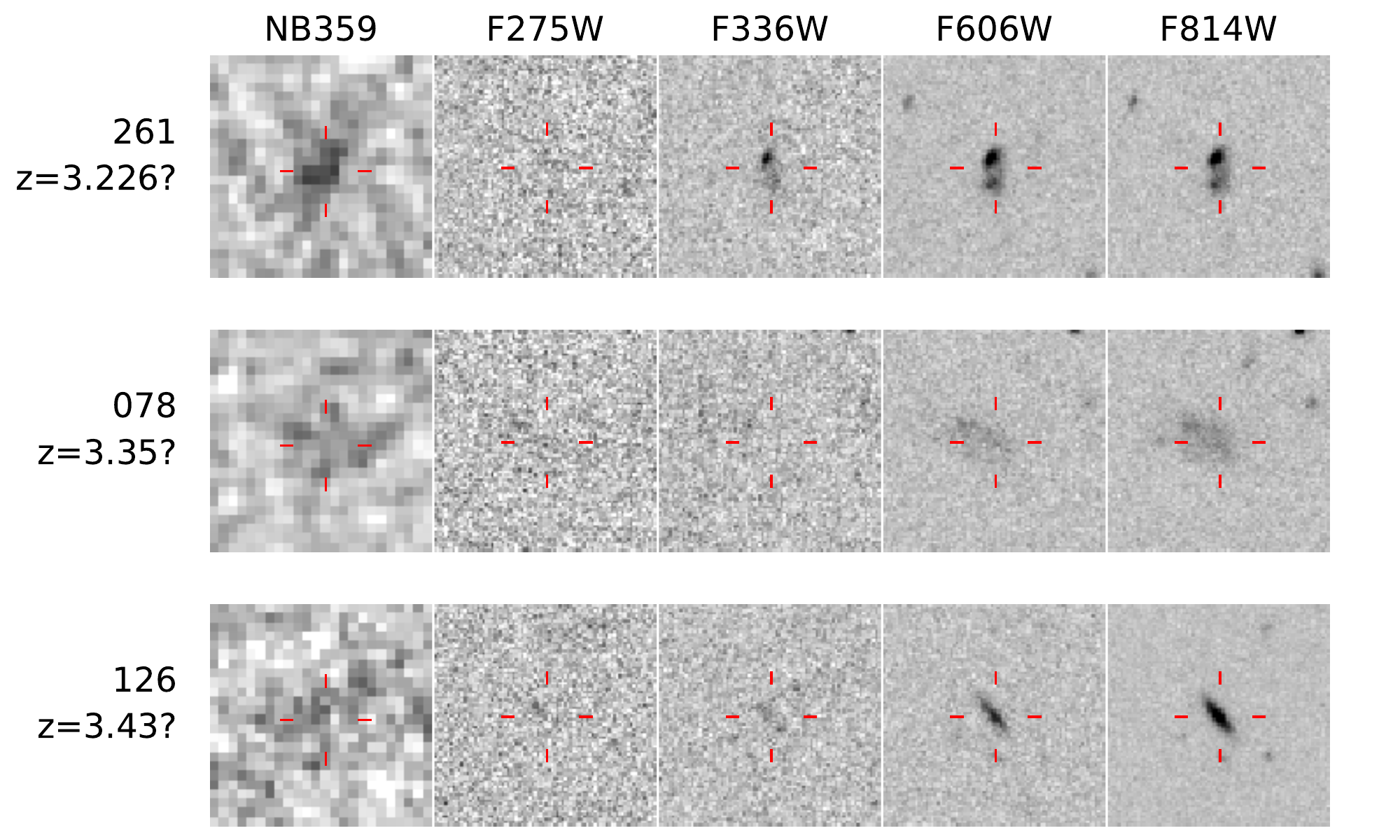}
 \caption{Subaru / Suprime-Cam \flyc and {\textit{HST}}
 WFC3/UVIS and ACS images of \flyc detected sources in the
 catalogues with redshifts between 3.1 and 3.5 but their redshifts are
 flagged as uncertain.}
 \label{fig:montage_unc}
\end{figure}

In the compilation of galaxies with spectroscopic redshifts, there are
several redshifts flagged as uncertain in the original catalog and with
no additional spectroscopic observation to confirm these redshifts, or
objects with conflicting redshift inferences by different authors. There
are seven such sources in the redshift range $3.16 < z < 3.43$, and
three objects among them have $\geq$3$\sigma$ detection in
\textit{NB359}.
In Fig.~\ref{fig:montage_unc} their postage stamp images are shown, and
Table~\ref{tab:unc_lyc_properties} provides a list of these objects. 
Although object 261 (MOSDEF~13286) is listed with a redshift quality
flag 7 (based on multiple emission lines or robust absorption-line
redshift) in the MOSDEF catalog, 3D-HST grism spectrum in
\citep{Pirzkal2013} shows multiple emission lines such as H$\beta$,
[O{\sc ii}] and [O{\sc iii}] with a best estimate redshift of 2.01. In
Fig.~\ref{fig:montage_unc} there are two separated components in the
{\textit{HST}} images and they might be two independent objects at
$z=2.0$ and $3.2$, although in the F336W image both components appear to
be detected. 
Without further spectroscopy we cannot conclude whether the \flyc signal
comes from the object at $z=3.2$, and we treat this object as an object
with uncertain redshift. 
For the other two objects in Table~\ref{tab:unc_lyc_properties},
\citet{Lowenthal1997} and \citet{Dawson2001} noted that the quality of
their optical spectra is too low to conclude their redshifts
confidently. Deeper spectroscopy would be necessary to determine whether
they are LyC emitters or not. In the following analyses we exclude all
of these objects with uncertain redshifts.

\citet{Grazian2017} used LBT / LBC {\textit{U}}-band imaging data to
search for LyC leakage from galaxies in redshift range $3.27<z<3.40$ in
multiple fields, including the GOODS-N. They reported a detection of one
galaxy in the GOODS-N at $z=3.371$ with a careful note of possible
contamination by a nearby source. In our \flyc image the signal from
this object is $2.7\sigma$ with a $1\farcs2$ diameter aperture, and the
object is not included in our list of LyC candidates.

\begin{table*}
 \centering
\caption{List of \flyc detected sources among objects in the
 catalogues with redshifts between 3.1 and 3.5 but with redshifts 
 flagged as uncertain.
}
\label{tab:unc_lyc_properties}
\begin{tabular}{clcclccl}
\hline
    & Designation  & R.A.(J2000) & Decl.(J2000) & Redshift & $m_{\mathrm{NB359}}^a$ & Ref.$^b$ & Note \\
\hline
261 & MOSDEF~13286    & 189.21385 & 62.20760 & 3.226? & $26.53\pm0.20$ & 2 & $z=2.01$ in \citet{Pirzkal2013} \\
078 & hd2\_0853\_0319 & 189.19540 & 62.22436 & 3.35?  & $27.02\pm0.26$ & 4 & \\
126 & F36568-1353     & 189.23692 & 62.23152 & 3.43?  & $27.19\pm0.28$ & 5 & \\
\hline
\end{tabular}
\begin{flushleft}
$^a$: Magnitudes in \flyc are measured with a $1\farcs6$ diameter aperture.
$^b$: Redshift references: (2) \citet{Kriek2015}, (4) \citet{Lowenthal1997},
(5) \citet{Dawson2001}.
\end{flushleft}
\end{table*}

\subsection{Near-infrared spectroscopy}
\label{subsec:nirspec}

\begin{figure*}
 \includegraphics[width=\columnwidth]{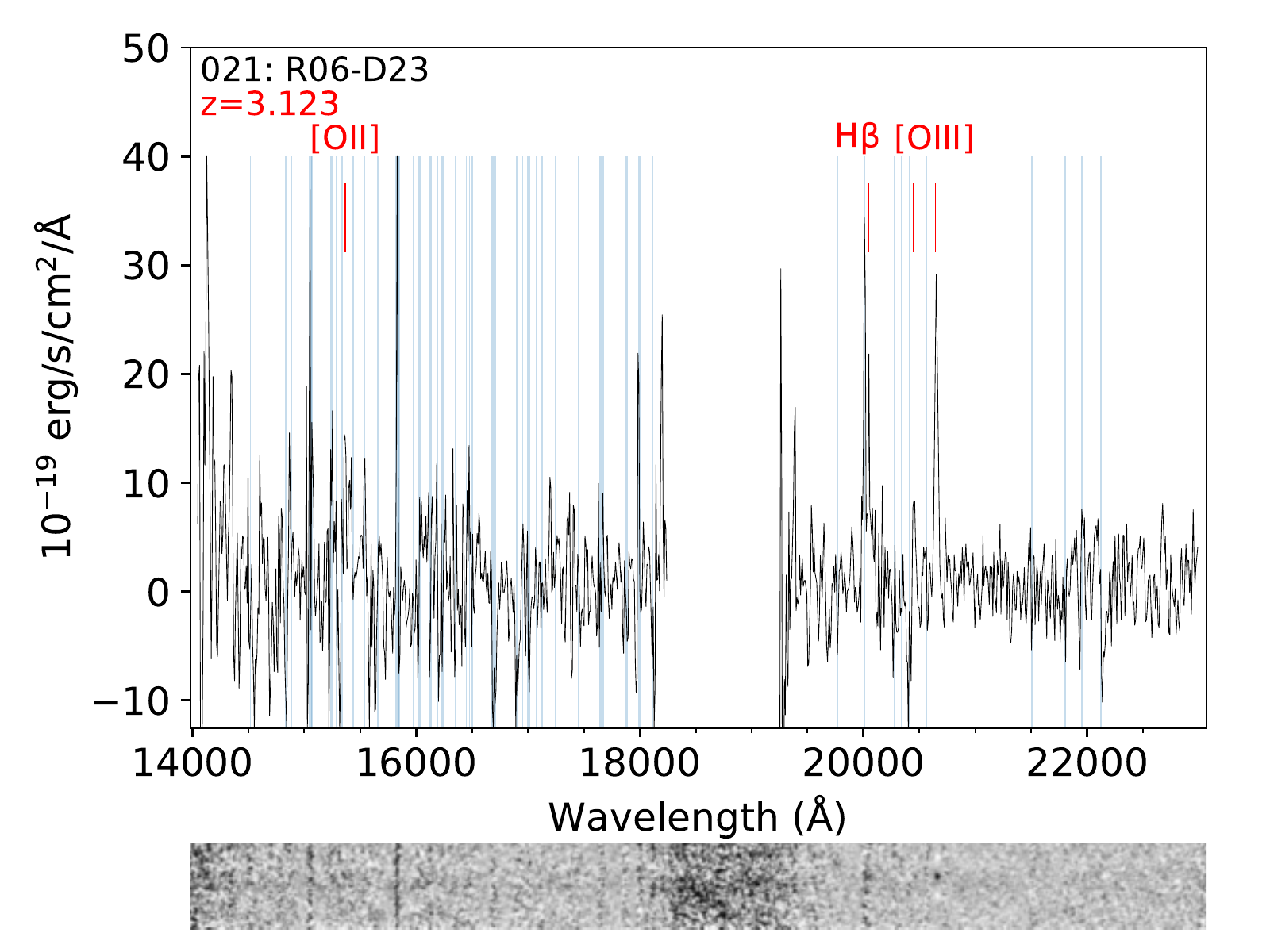}
 \includegraphics[width=\columnwidth]{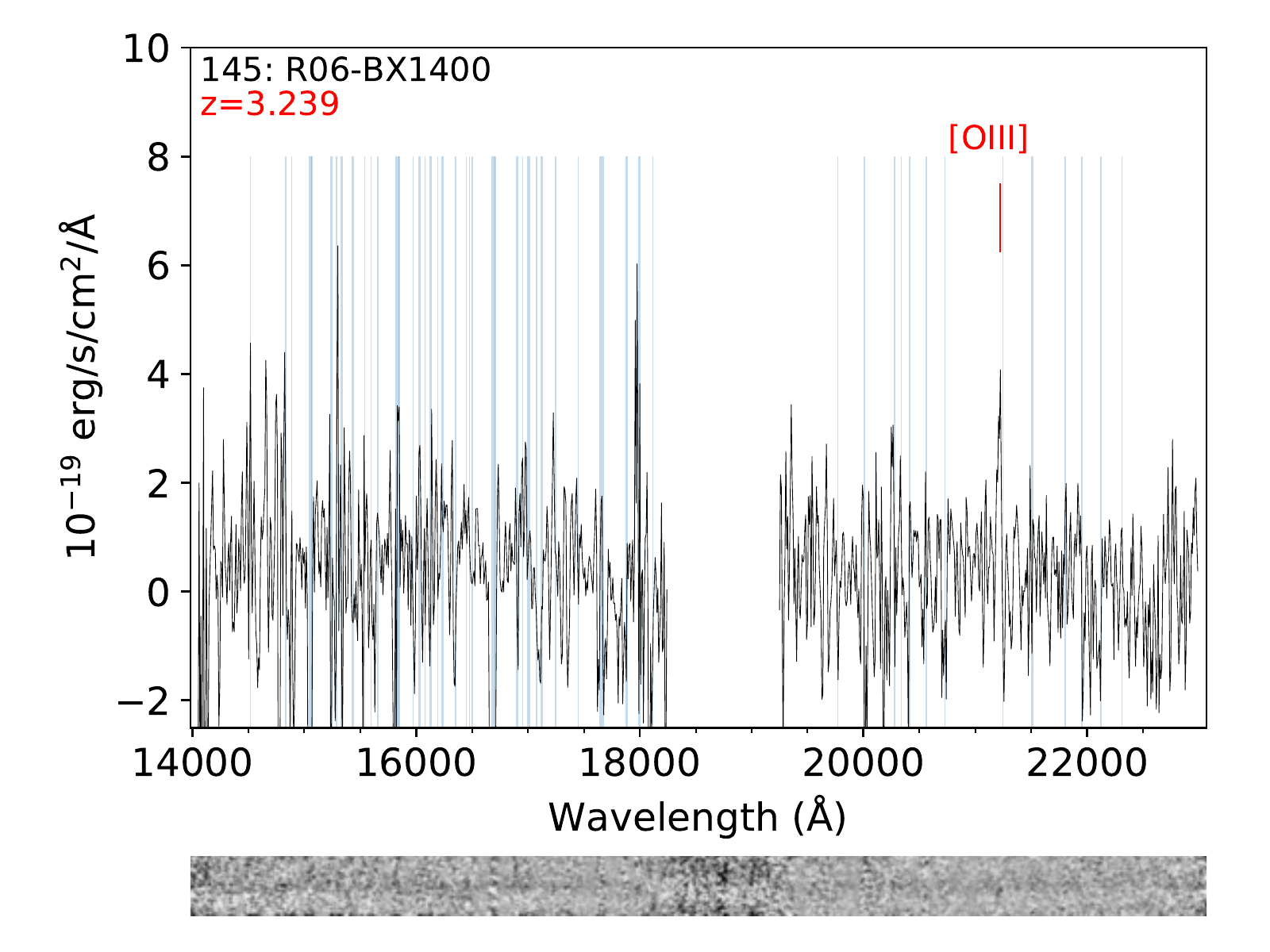}
 \includegraphics[width=\columnwidth]{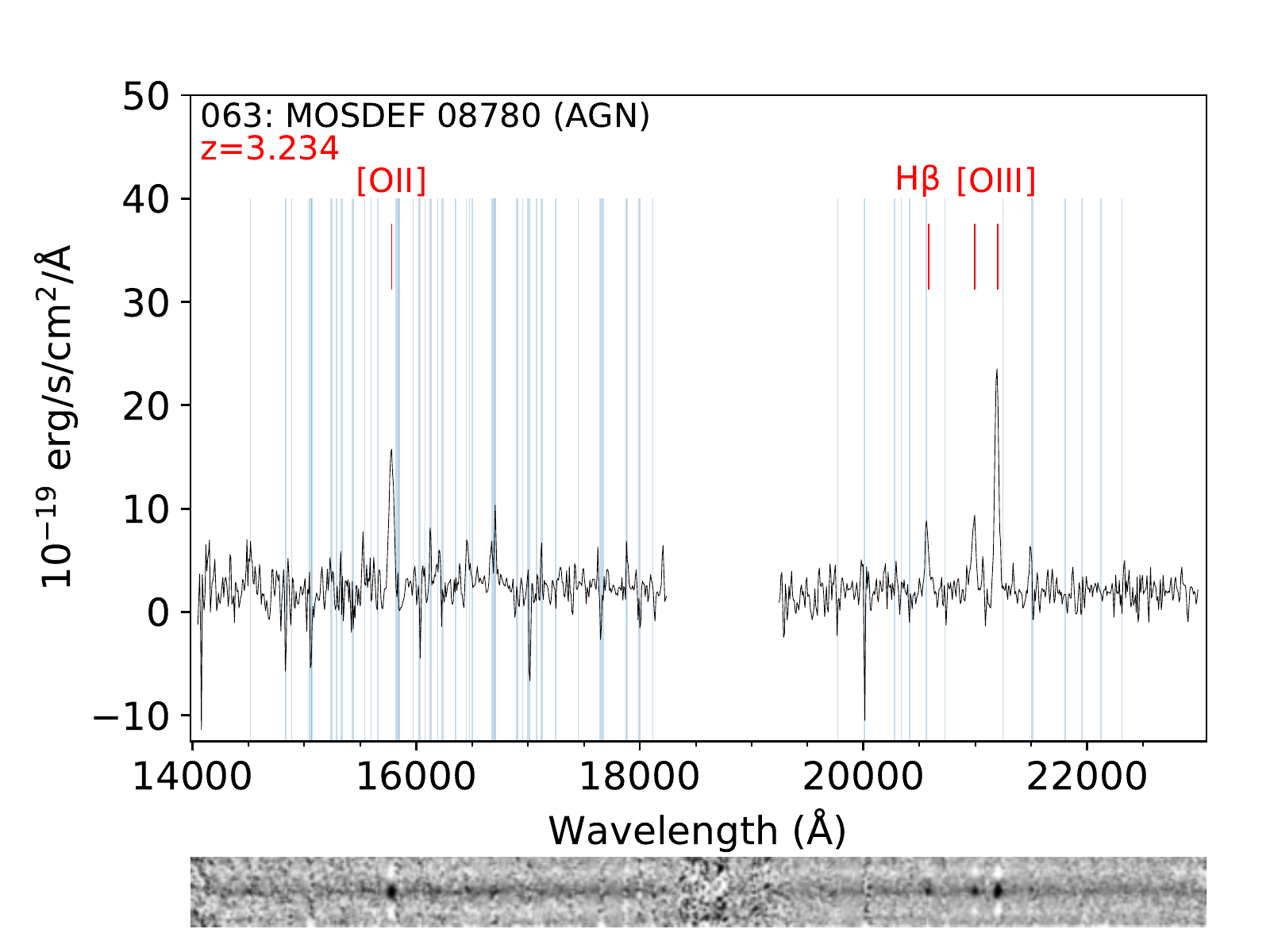}
 \includegraphics[width=\columnwidth]{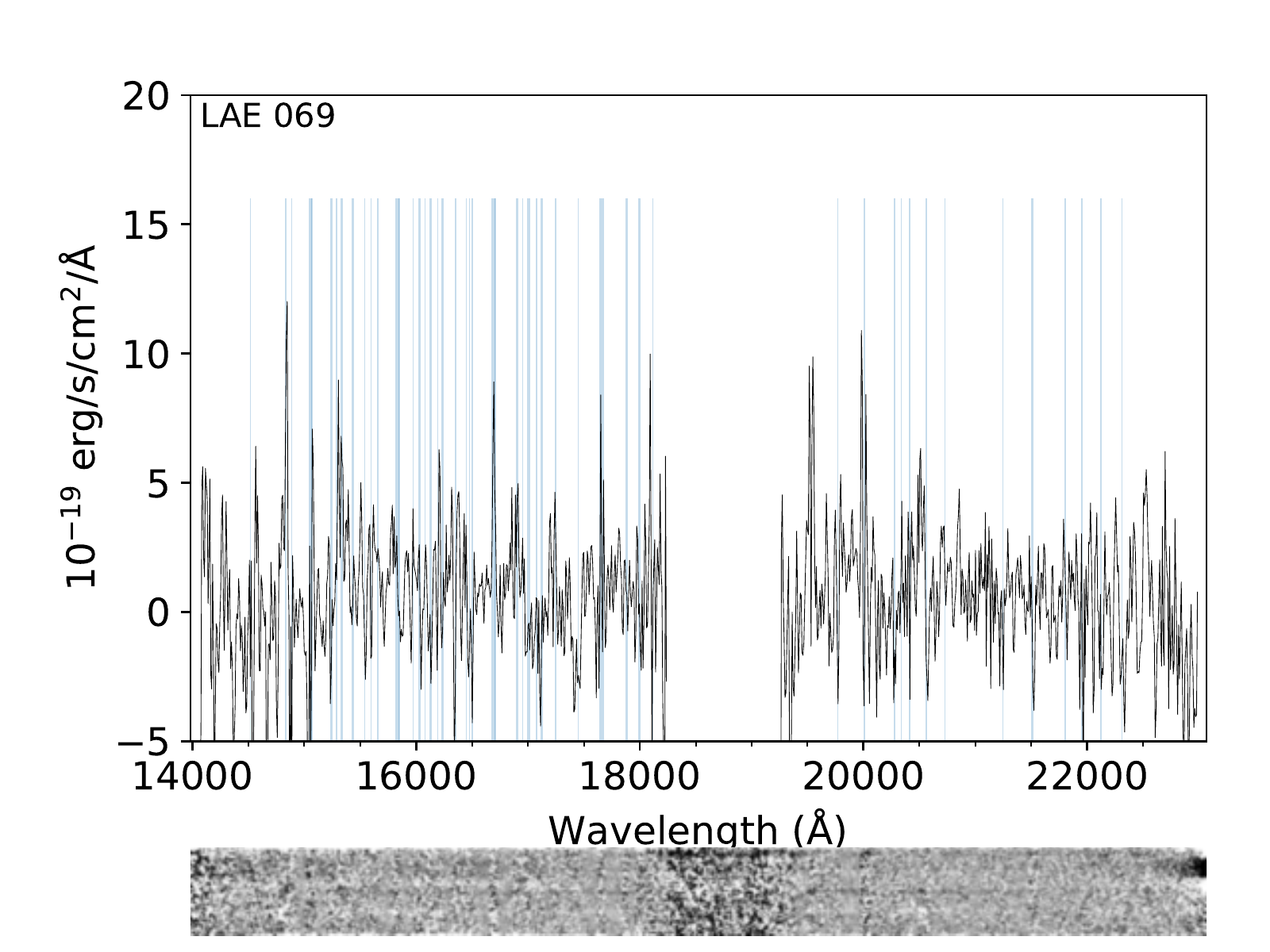}
 \caption{MOIRCS {\textit{H}} and {\textit{K}}-band spectra of four
 objects among galaxies with \flyc detection. 021 and 145 are
 star-forming galaxies with known spectroscopic redshifts, and 063 is a
 galaxy at $z=3.234$ hosting an AGN. Objects 145 and 063 are reportedly
 contaminated by foreground galaxies at $z\sim 0.5$ \citep{Jones2018}
 but no sign of contamination is identified in this wavelength
 range. There is no feature found in the spectrum of LAE 069, which is
 expected to be at around $z\sim 3.1$. Shaded areas
 indicate wavelength ranges with atmospheric OH lines. Emission line
 features are indicated with red vertical lines. Two-dimensional
 reduced spectral images are shown in the bottom of the panels.}
 \label{fig:moircs_zsp}
\end{figure*}

In our MOIRCS spectroscopy, among the objects detected in the
\flyc image, two objects in the SFG sample, one object in the AGN
sample, and one object in the LAE sample are observed. 
Fig.~\ref{fig:moircs_zsp} shows their {\textit{H}} and {\textit{K}}-band
spectra. 
For object 021 (R06-D23) our near-infrared spectrum confirms the
reported redshift $z=3.123$ by \citet{Reddy2006}.
\citet{Jones2018} reported contamination by a foreground source
at $z \sim 0.5$ for both object 145 (R06-BX1400) and object 063
(MOSDEF~08780) through their optical spectroscopy.
In the {\textit{H}} and {\textit{K}}-band wavelength range there is no
strong feature expected from an object at such redshift, and we do not
identify any sign of foreground sources for these two objects.
The spectrum of LAE~069, which is among the sample selected with an
excess in the \textit{NB497} image, does not show any feature other than
weak continuum, and we could not confirm its redshift.

\begin{figure}
 \includegraphics[width=\columnwidth]{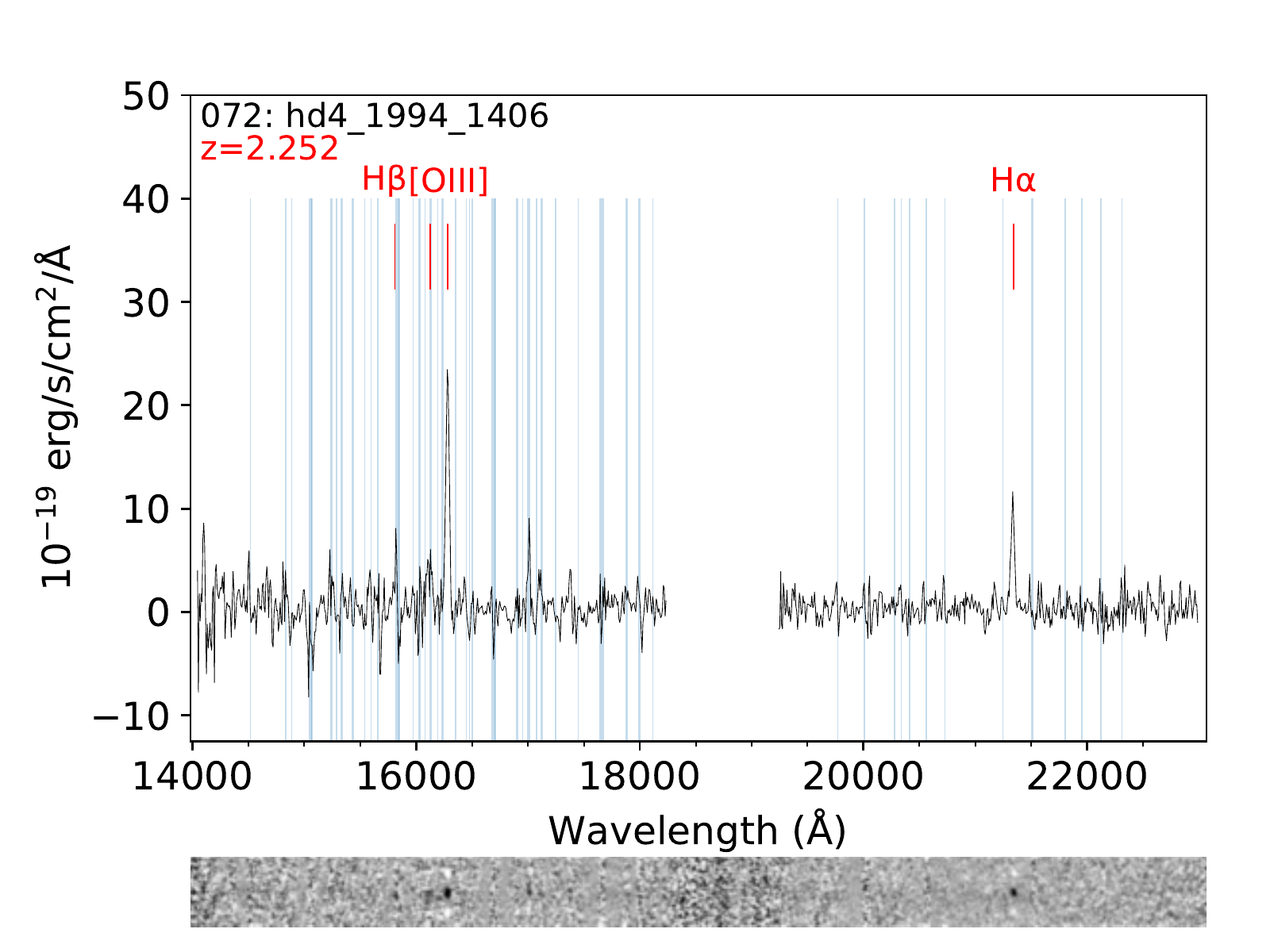}
 \caption{MOIRCS {\textit{H}} and {\textit{K}}-band spectrum of
 object 072, hd4\_1994\_1406. In \citet{Lowenthal1997} its redshift was
 tentatively estimated to be 3.63, but this near-infrared spectrum
 determines its redshift to be 2.252.
}
 \label{fig:moircs_072}
\end{figure}

In Table~\ref{tab:moircs} we summarize the redshifts of observed targets
for which multiple and conflicting redshift estimates were available in
the literature or our redshift determinations are new.
Object 072 (hd4\_1994\_1406 in \citet{Lowenthal1997}) has a redshift
estimate $z=3.63$ with uncertainty, and our near-infrared spectroscopy
determined its redshift as $z=2.252$ with identification of H$\beta$,
[O{\sc iii}] and H$\alpha$ emission lines. 
Fig.~\ref{fig:moircs_072} shows the MOIRCS spectrum of the object.
Although this object is detected in \textit{NB359}, with this lower
redshift \flyc does not trace ionizing radiation.
Also, with our spectroscopy we confirmed redshifts of three objects in
the MOSDEF survey, namely MOSDEF~17204, MOSDEF~18436 and MOSDEF~12302,
which have conflicting redshift entries in the literature. The lower
redshifts in MOSDEF are confirmed, and therefore for these objects \flyc
does not trace ionizing radiation. Among \textit{NB497}-selected LAE 
candidates not detected in \textit{NB359}, we list three redshift
identifications with the [O{\sc iii}]$ \lambda$5007 line, although for
two of them the line detections are marginal. 

\begin{table*}
 \centering
\caption{Notable redshift information from the result of MOIRCS spectroscopy.
}
\label{tab:moircs}
\begin{tabular}{lcclll}
\hline
 Designation  & R.A.(J2000) & Decl.(J2000) & Redshift & Lines & Notes\\
\hline
MOSDEF~17204 & 189.16877 & 62.22633 & 1.490 & H$\alpha$ & F36405$-$1334:$z=3.826$ in \citet{Dawson2001}, \\
             &           &          &       &           & $z=1.4879$ in \citet{Kriek2015} \\
MOSDEF~18436 & 189.17022 & 62.23287 & 2.993 & [OIII]4959,5007 & R06-MD43:$z=3.0870$ in \citet{Reddy2006}, \\
             &           &          &       &                 & $z=2.9894$ in \citet{Kriek2015} \\
hd4\_1994\_1406 & 189.18885 & 62.19403 & 2.252 & H$\beta$,[OIII]5007,H$\alpha$ & $z=3.63$ in \citet{Lowenthal1997} \\
MOSDEF~12302 & 189.30953 & 62.20236 & 2.275 & H$\beta$,[OIII]5007,H$\alpha$ & AGN; $z=3.1569$ in \citet{Barger2008}, \\
             &           &          &       &                 & $z=2.2756$ in \citet{Kriek2015} \\
LAE082 & 188.84854 & 62.16625 & 3.083  & [OIII]5007 & \\
LAE097 & 188.86138 & 62.18704 & 3.091? & [OIII]5007? & \\
LAE091 & 189.18979 & 62.17562 & 3.085? & [OIII]5007? & \\
\hline
\end{tabular}
\end{table*}

\section{Properties of LyC Candidates}
\label{sec:lyc_properties}

Among 103 SFGs with secure spectroscopic redshifts within a range
$3.06<z<3.5$, three galaxies are detected with \flyc at $\geq$3$\sigma$
level. One object among them (running number 145, R06-BX1400) is found
to be contaminated by a foreground object by \cite{Jones2018}, and thus
there are two SFGs with possibly genuine LyC emission.
Among 157 $z\sim 3.1$ LAE candidates, there are five objects detected
with \textit{NB359}. 
None of these LAE
candidates has a spectroscopically confirmed redshift, although 
by spectroscopic observations of the LAE candidates in the
SSA22 field selected with the same \flya\ filter it has been reported
that the contamination rate is as small as $\sim$1\% \citep{Yamada2012}.
Here we examine the properties of the remaining two star-forming LyC
emitter candidates, as well as those of the \flyc detected LAE
candidates.

\subsection{Spectral energy distribution}
\label{subsec:lyc_SED}

\begin{figure}
 \includegraphics[width=\columnwidth]{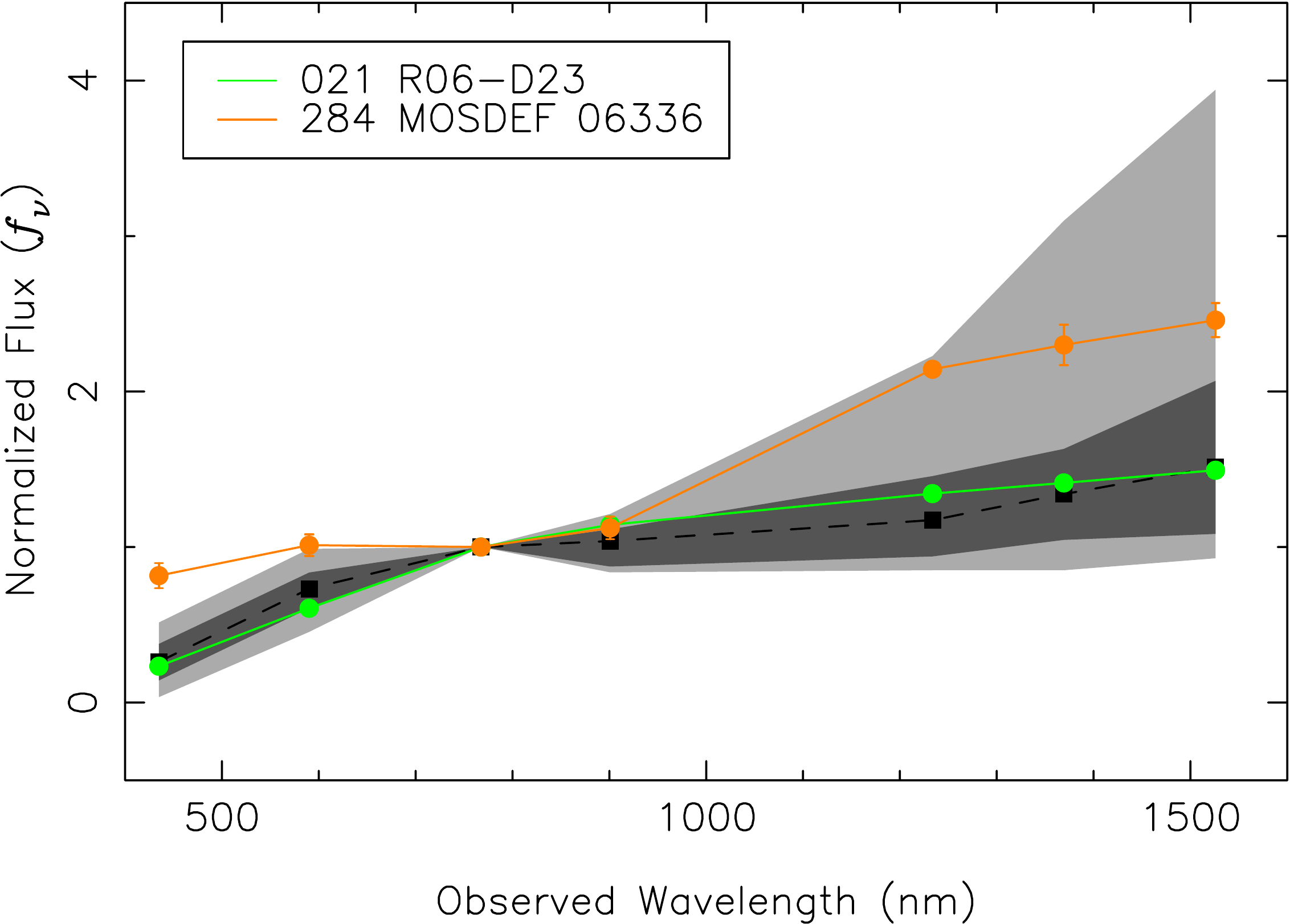}
 \caption{Normalized SEDs in 400 -- 1600 nm (rest-frame UV)
 wavelength range for the SFG sample galaxies. Flux densities are
 normalized with that
 in $I_c$-band. Dark and light shaded areas show the 68\% and 95\%
 ranges of the SEDs of the 102 SFGs. The black points connected with
 dashed lines are the median values. The orange and green points show
 the SEDs of object 021 (R06-D23) and object 284 (MOSDEF~06336),
 respectively.
 }
 \label{fig:optical_SED}
\end{figure}

First we examine the SEDs of the two LyC emitting SFG candidates.
Fig.\,\ref{fig:optical_SED} shows their flux densities in
{\textit{HST}}/ACS F435W, F606W, F850LP, WFC3/IR F125W, F140W, F160W,
normalized by the flux density in F775W. 
These flux densities are taken from the 3D-HST photometric catalog
\citep{Skelton2014}.
The flux density distribution in each filter for the entire SFG sample 
(excluding one object which does not have any entry in the catalog by
\citet{Skelton2014}) is indicated by  the shaded areas which contain
68\% and 95\% of the sample galaxies. 
The redshift range of the sample galaxies spans from $z=3.06$ to 3.5,
and the corresponding rest-frame wavelength range shown in this figure
is from $\sim$1000\AA\ to $\sim$4000\AA. Although the SED of object 021
(R06-D23) is within a 68\%-ile area for the entire SFG sample, the SED
of object 284 (MOSDEF~06336) is peculiar; its rest-frame far-UV SED
(traced with F435W,  F606W, and F775W) is very flat, while there is a
spectral break between F850LP and F125W. Because this object is reported
to be at $z=3.412$, F435W traces rest-frame wavelength shorter than
Ly$\alpha$, and such a flat SED at far-UV needs optically thin IGM or an
extremely blue SED. However, the red SED at longer wavelengths seems to
be in conflict with the latter option.

\begin{figure*}
 \includegraphics[width=0.9\columnwidth]{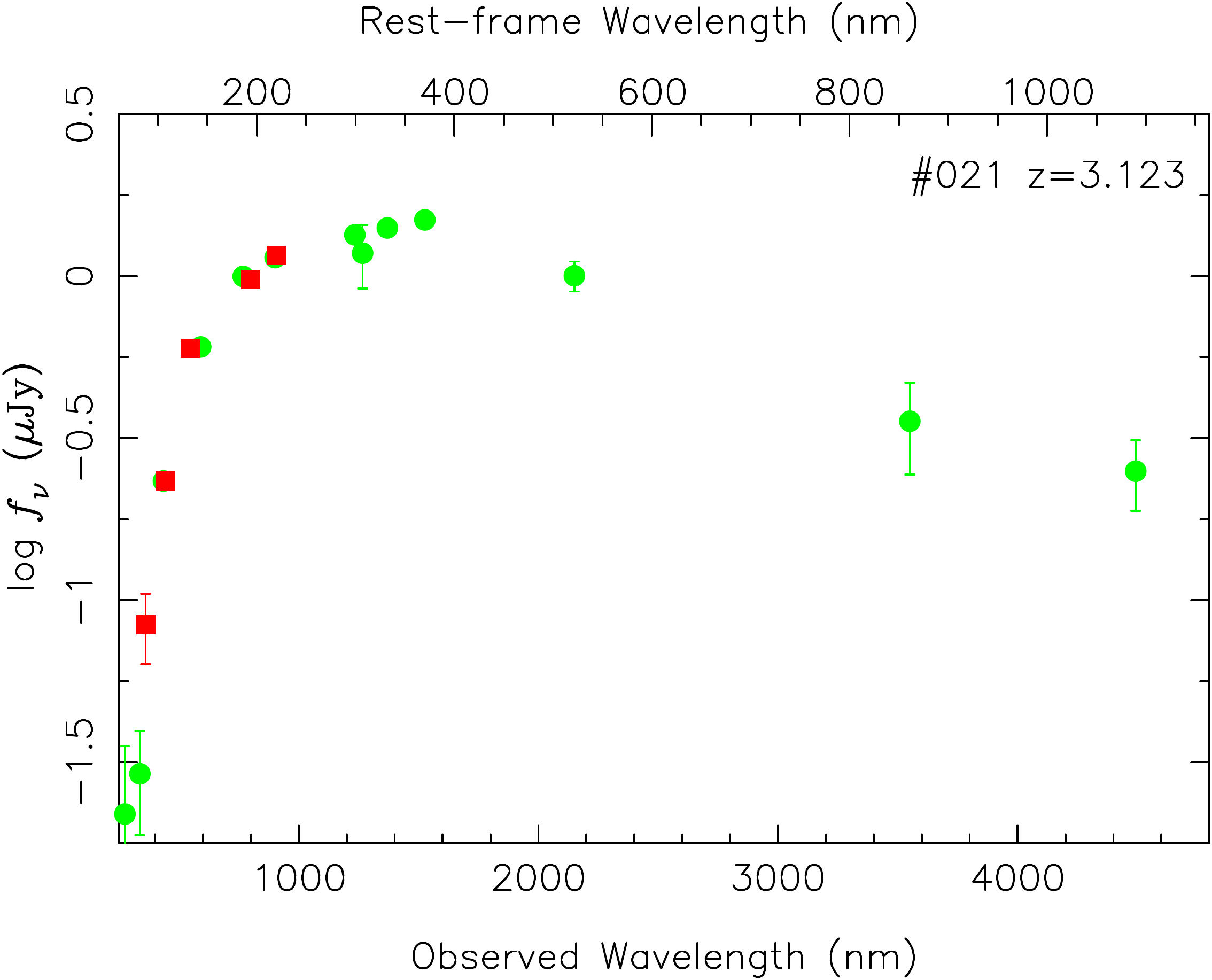}
 \hspace{0.1\columnwidth}
 \includegraphics[width=0.9\columnwidth]{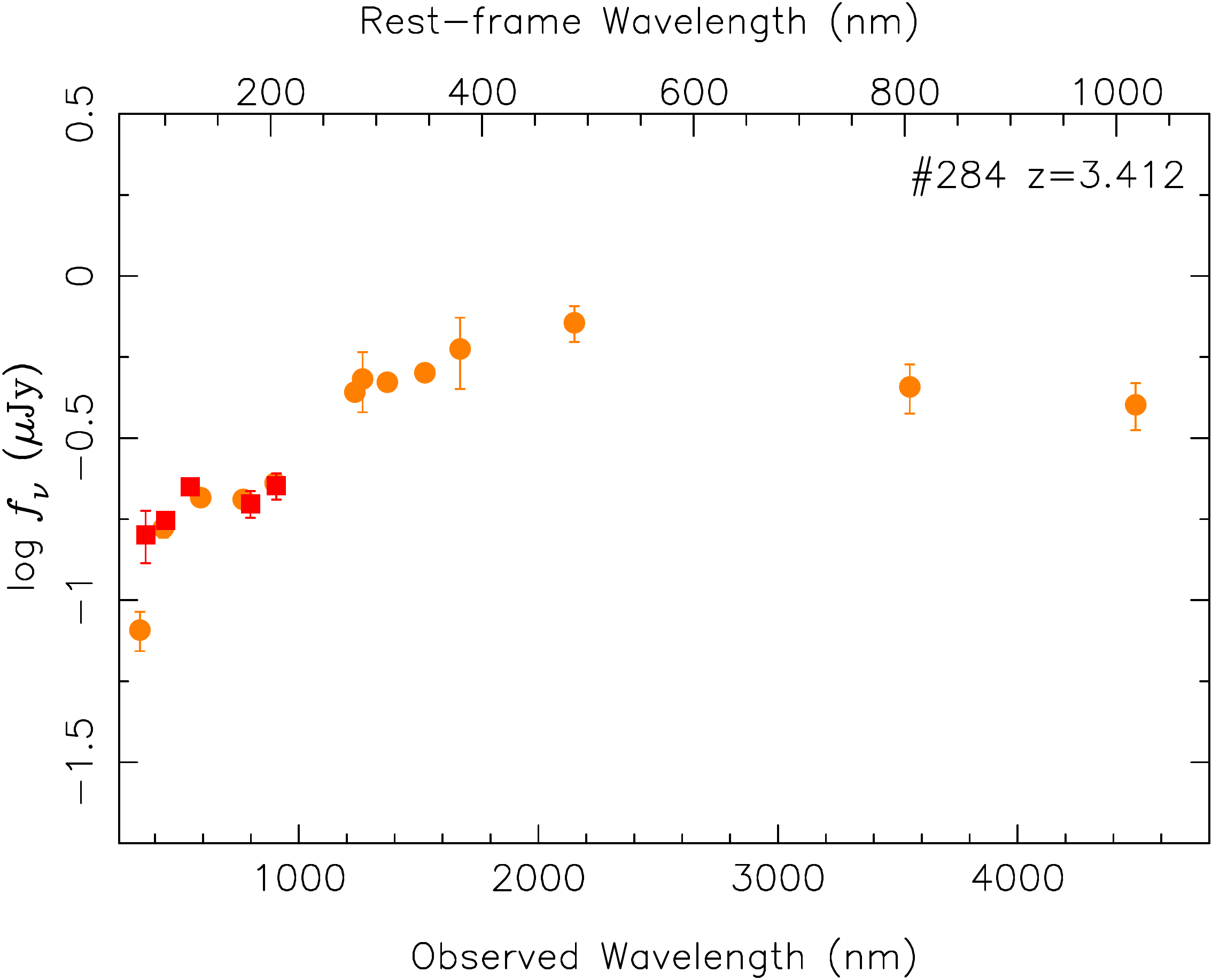}
 \caption{SEDs of two star-forming SFG LyC candidates.
 Circles are flux densities taken from the HDUV catalog
 \citep{Oesch2018} for F275W and F336W data points, and from the 3D-HST
 photometric catalog \citep{Skelton2014} for the other points.
 Squares points in red stand for Suprime-Cam photometry.
 }
 \label{fig:SED_lyc}
\end{figure*}

Fig.\,\ref{fig:SED_lyc} shows the SEDs of the two LyC emitting SFG
candidates in broader wavelength range. Data points for near-IR
photometry from MOIRCS Deep Survey \citep{Kajisawa2011} and Spitzer IRAC
Channel 1 and 2 photometry from SEDS \citep{Ashby2013}, taken from the
photometric catalog by \citet{Skelton2014}, as well as Suprime-Cam
photometric data points are plotted.
In Suprime-Cam photometry a $1\farcs6$ diameter aperture is used, and
the factor to scale the aperture flux to the 3D-HST total flux is
calculated by averaging the flux ratios between the 3D-HST total flux
and the Suprime-Cam aperture flux in {\textit{B}}, {\textit{$I_c$}}, and
{\textit{z}}-band, and the factor is applied to the flux densities in
all Suprime-Cam bands.

Again, the SED of object 284 is puzzling. In addition to the spectral
break around 1$\mu$m observed wavelength, the flux ratio between \flyc
and $I_c$ is 0.80. This means that if this object is at $z=3.412$, the
observed \flycfuv\, without correction for IGM attenuation is close to
unity.
If we correct the flux ratio assuming the average IGM attenuation at the
redshift ($\langle t_{\mathrm{IGM}} \rangle=0.15$ for \textit{NB359}),
the LyC/UV flux ratio exceeds 6. However, as shown in
Fig.~\ref{fig:igm_trans}, a significant fraction of sightlines toward
objects at $z\sim 3.4$ has IGM transmission as high as $\sim$0.5
(according to the Monte Carlo simulation described in
Section~\ref{sec:igm_trans}, $\sim$18\% of sightlines have an IGM
transmission higher than 0.4 for \flyc).
If the IGM toward object 284 has such a low opacity, the IGM-corrected
LyC/UV flux ratio would be $\approx2$.
Population synthesis models of metal-poor, young stellar populations
predict relatively high \flycfuv\, values. 
For example, BPASS version 2.1 \citep{Eldridge2017} with a constant star
formation rate, metallicity $Z=0.001$ or lower, an IMF slope $-2.35$
from 0.5 to 300 $M_{\sun}$, without dust attenuation gives 
$0.3 \lesssim f_{\mathrm{LyC}}/f_{\mathrm{UV}} \lesssim 0.9$ 
for a galaxy with age 1 to 100 Myrs.
Although even with such metal-poor stellar populations \flycfuv\,
larger than unity is difficult to be explained, \citet{Inoue2010} showed
that an SED with escaping nebular bound-free LyC could make a LyC `bump'
just below the Lyman limit and intrinsic \flycfuv $>$1
\citep[see also detailed discussions in][]{Inoue2011}.
Therefore, although the SED of object 284 appears to be peculiar, we
cannot rule out the possibility that this object is a genuine LyC
emitter at $z\sim 3.4$. 
We further discuss the reality of the LyC detection of these two SFGs in
the next subsection.

\subsection{Reality of LyC detections}
\label{subsec:lyc_reality}

\subsubsection{Object 021: R06-D23}

Object 021 (R06-D23) is unresolved in the  \textit{HST}/ACS images,
which means that the FWHM is less than $0\farcs1 \sim 0.8$ kpc at
$z=3.12$. 
We confirmed its redshift with near-infrared spectroscopy
(Section~\ref{subsec:nirspec}). The optical spectrum of the object taken
with Keck/LRIS (C. Steidel, private communication) shows Ly$\alpha$
emission with line width of $\approx$700 km\,s$^{-1}$ and rest-frame UV
absorption lines such as O{\sc i}, C{\sc ii} which give its redshift
$z=3.123$, and there is no sign of broad-line AGN. No counterpart is
found in the Chandra 2Ms X-ray catalogs \citep{Alexander2003, Xue2016}.
There is no indication of a foreground source in the \textit{HST} images
and spectra. 
Although weak AGN may contribute to its LyC emission, based on these
observations we consider that this object is a real LyC emitting SFG at
$z=3.123$. 
This object is also identified as a star-forming LAE GN-NB5-5878 with
Ly$\alpha$ redshift of 3.129 in \citet{Sobral2018}.

\subsubsection{Object 284: MOSDEF~06336}

The situation for object 284 (MOSDEF~06336) is more complicated. Its
\textit{HST}/ACS images show faint nearby sources 
(see Fig.\,\ref{fig:montage_SFGAGN}), which may be physically associated
with the object or sources in the background / foreground.
The position of \flyc and F336W images align with the brightest
component. As discussed in Section~\ref{subsec:lyc_SED}, the SED of the
brightest component is difficult to be explained with a source redshift
of $z=3.4$. A quick photometric redshift estimate using EAZY
\citep{Brammer2008} with default templates returns
$z_{\mathrm{phot}}=2.02\pm0.07$ as the maximum probability photometric
redshift. 
On the other hand, MOSDEF $K$-band spectrum of the object (N. Reddy,
private communication) shows [O{\sc iii}] $\lambda\lambda$4959, 5007
and H$\beta$ which give its systemic redshift at $z=3.412$.
The slit configuration of the MOSDEF spectra is such that a fainter
nearby source to the south east may partially be falling in the slit,
merging with the spectrum of object 284.
The spectroscopic redshift of this nearby source is unknown 
but it has ID 6381 with a photometric redshift of $z\sim 0.5$ in the
3D-HST v4.1 photometric catalog \citep{Skelton2014}.
No major emission line is expected to be observed for a galaxy at $z
\sim 0.5$ in the near-infrared wavelengths.
The $Ks$-band flux density of the object from the MOIRCS Deep Survey
\citep{Kajisawa2011} is 0.29 $\mu$Jy.
On the other hand, the total flux of [O{\sc iii}]+H$\beta$ is 
$4.6 \times 10^{-17}$ erg\,s$^{-1}$\,cm$^{-2}$, 
and it corresponds to 0.23 $\mu$Jy when diluted to $Ks$-band. 
Because the fainter nearby source only partially falls into the slit, it
is unlikely that it can fully account for the 
[O{\sc iii}] and H$\beta$ emissions. Therefore, we conclude that these
emission lines at $z=3.412$ originate from the brightest object where
\flyc and F336W images are aligned.
Although the SED of the object is difficult to interpret, without any 
strong evidence for foreground contamination we retain the object as a
candidate LyC emitting galaxy, but we keep in mind that this object is
not a highly reliable LyC 
candidate\footnote{Flux density measured with \flyc for object 284 is
about a factor of 2 higher than in F336W. This may be explained by the
difference in band widths of these filters and a variation of the IGM
attenuation by H{\sc i} clouds at different redshifts (see
Fig.\,\ref{fig:igm_trans}).}.

\subsubsection{\flyc-detected LAEs}
\label{sec:reality_LAE}

Among the five LAE candidates with a $\geq$3$\sigma$ detection in \flyc, 
flux densities in \flyc of both LAE019 and LAE033 are 3.2$\sigma$, which
is smaller than the values for the other LAEs ($>3.8\sigma$), and
their postage stamp images shown in Fig.~\ref{fig:montage_LAE} are less
convincing. In order to test the significance of the flux signal in \flyc
for these objects, we split the reduced individual observed frames into
two groups, stacked them separately to produce two independent mosaiced
images with half the total integration time, and measured flux densities
of the five objects with a $1\farcs2$ diameter aperture.
While the measured flux densities of the other
three objects are $\gtrsim2\sigma$ in both mosaiced images of
half the total integration time, the flux densities are  1.0$\sigma$ and
2.2$\sigma$ for LAE019, and 1.5$\sigma$ and 1.4$\sigma$ for LAE033.
The flux measurements in the mosaiced images are consistent with each
other within errors estimated from the measured background fluctuation
of the images, and this test excludes a possibility that these
detections are based on a spike noise in a single frame. We thus retain
LAE019 and LAE033 as candidates of LyC detection, although the
significance of the signals for these objects is smaller than for the
rest.

As described in Section~\ref{subsec:nbphot}, in two among the five
candidates the \flyc signal appears to be associated with an object
spatially separated from the \flya\ detection, and we consider that
these two objects are likely to be contaminated by foreground sources. 

The \flycfuv\ flux ratios of three LAEs exceed unity
(Table~\ref{tab:lyc_properties}).
Although such high \flycfuv\ is possible as discussed in
Section~\ref{subsec:lyc_SED}, an overlap of a foreground source can also
explain such flux ratios. 
Indeed, two among the three are LAE053 and LAE137 which have
large spatial offsets between \flyc and \flya (0\farcs81 and 0\farcs59,
respectively; see Table~\ref{tab:lyc_properties}). Although the spatial
offset for the other object LAE033 is smaller (0\farcs31) than these two
objects, we also treat the object as one possibly contaminated by a
foreground object based on its high \flycfuv (1.30). 
Thus among the five LAEs with $>3\sigma$ \flyc signal, three are
considered to be possibly contaminated by foreground objects.
In Section~\ref{subsec:foreground} we examine
statistical probabilities of foreground contamination, and there is a
non-negligible likelihood that all of these \flyc detected LAE
candidates are contaminated by foreground sources. Deep optical /
near-infrared spectroscopy and / or sensitive imaging with high spatial
resolution are required to further assess the reality of these LyC
emitting LAE candidates.

There must be a selection bias that we preferentially select objects
with high \flycfuv\ for objects with faint rest-frame UV continuum, due
to the sensitivity limit of our \flyc imaging observations. However, as
we will see in Section~\ref{sec:comparison_ssa22}, we have not detected
sources with high \flycfuv\ in UV-bright sources, and there would be a
correlation between UV luminosity and \flycfuv\ (and hence the escape
fraction of LyC; see \citet{Fletcher2018}).

\subsection{Ly$\alpha$ equivalent widths and LyC escape}

\begin{figure}
 \includegraphics[width=\columnwidth]{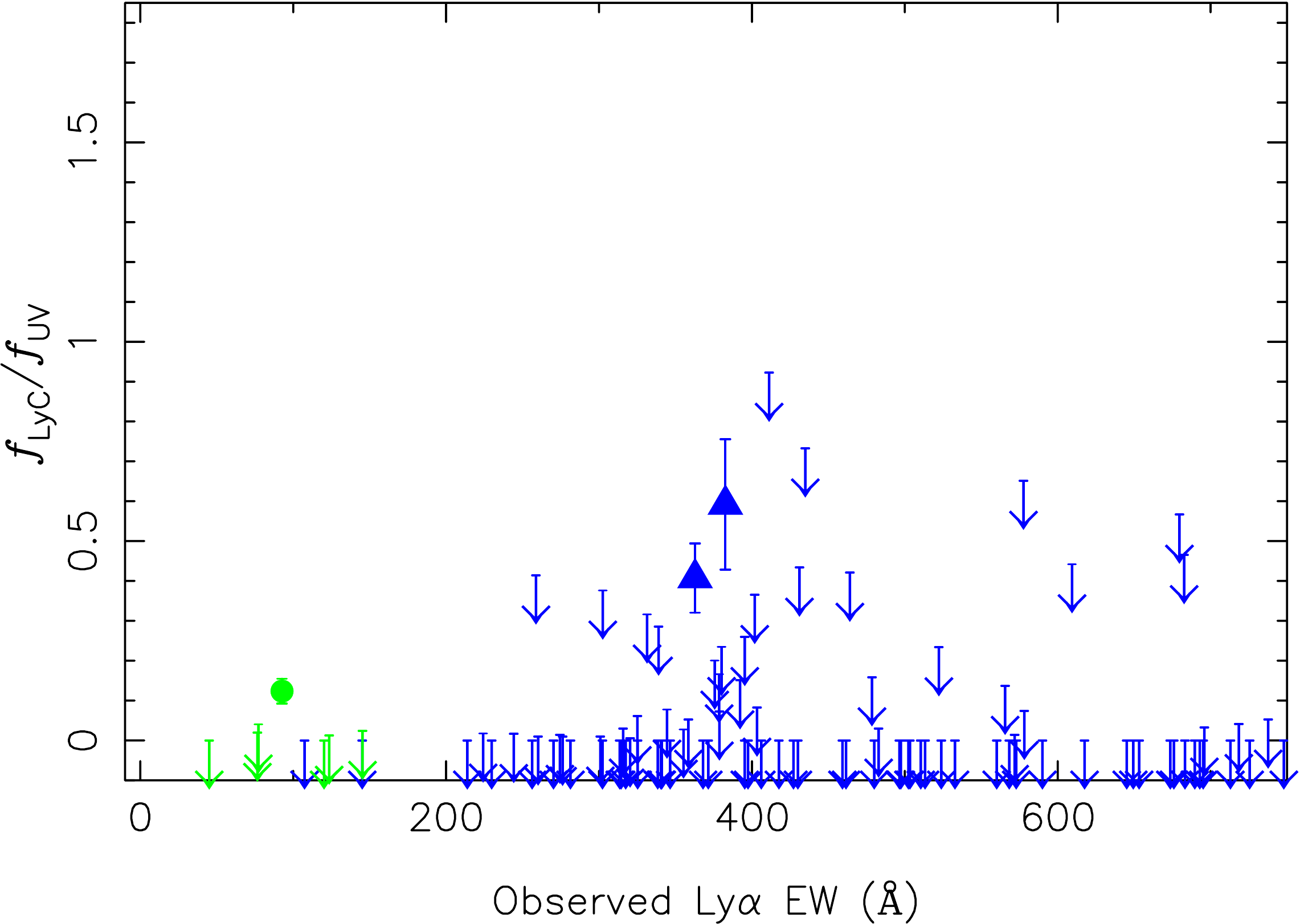}
 \caption{
 Observed Ly$\alpha$ EWs and $f_{\mathrm{LyC}}/f_{\mathrm{UV}}$ for a
 subsample among SFGs at $3.06<z<3.126$ (green) and \flya-selected
 LAE candidates (blue).
 $f_{\mathrm{LyC}}/f_{\mathrm{UV}}$ ratios are measured with a
 $1\farcs6$ diameter aperture. Galaxies with \flyc flux measurements
 less than 3$\sigma$ are shown with arrows.
 }
 \label{fig:LyaEW_LyCUV}
\end{figure}

For galaxies at redshifts between 3.06 and 3.126 \flya\,
can be used to measure the Ly$\alpha$ flux density.
In addition to the LAE sample, there are seven galaxies in the SFG
sample which are within this redshift range. Among them, one object 
(021 R06-D23; see Table~\ref{tab:lyc_properties}) is detected in
\textit{NB359}.
In Fig.\,\ref{fig:LyaEW_LyCUV} the observed Ly$\alpha$ equivalent widths
(EWs) of LAEs and these SFGs are plotted against LyC to rest-frame
non-ionizing UV flux ratios.
Ly$\alpha$ EWs are calculated following \citet{Matsuda2009,
Micheva2017}, using a `{\textit{BV}}' image,
BV=((2{\textit{B}}+{\textit{V}})/3):
\[
EW(Ly_{\alpha})=\frac{(887+936)\times77\times(10^{0.4(BV-NB497)}-1)}{(887+936)-0.6\times77\times10^{0.4(BV-NB497)}}
\]
where $887$ \AA, $936$ \AA, and $77$ \AA~are the FWHMs of the
{\textit{B}}, {\textit{V}} and \textit{NB497}, respectively, and
{\textit{BV}} and \flya\ represent magnitudes in these bands.
In Fig.\,\ref{fig:LyaEW_LyCUV} objects with a \flyc signal
smaller than $3\sigma$ are plotted with downward arrows.

In \citet{Micheva2017}, a possible correlation between the Ly$\alpha$
EW and the LyC-to-UV flux ratio for $z\sim3.1$ LAEs and LBGs in the
SSA22 field has been pointed out. The trend seen in the GOODS-N is
qualitatively similar, although the number of \flyc detections in the
present sample is smaller.
\citet{Fletcher2018} also found a positive correlation between the
Ly$\alpha$ EW and the LyC escape fraction for their SSA22 sample.
As noted by these authors, this correlation may be a secondary relation
since both LyC-to-UV flux ratio (and LyC escape fraction which is
derived from LyC-to-UV flux ratio with corrections for IGM/CGM
attenuation and intrinsic flux ratio) and Ly$\alpha$ EW depend on the
non-ionizing UV flux density.
\citet{Steidel2018} argued that there is a tight correlation between the
covering fraction of the interstellar medium (ISM) and the
Ly$\alpha$ EW. The covering fraction and H{\sc i} column density of the
ISM were estimated through fitting spectral synthesis models with simple
geometries of the ISM by damping wings of Ly$\alpha$ and $\beta$ 
absorption lines, the depth of higher order Lyman series, and the
residual LyC flux of the composite spectra of their LBG sample
galaxies.
They suggested that the ISM geometry with `holes', through which both LyC
and a fraction of Ly$\alpha$ could escape, would explain the tight
correlation between the estimated covering fraction, the Ly$\alpha$ EW,
and the LyC escape fraction.

Such a correlation between Ly$\alpha$ EWs and the LyC escape fraction is
also reported for LyC emitting galaxies at lower redshifts
\citep{Verhamme2017}.
\citet{Verhamme2015} argued that in star-forming regions with a covering
fraction less than unity or density-bounded nebulae, optically thin to
LyC, there will be significant LyC leakage \citep{Zackrisson2013}, and
Ly$\alpha$ emission will have prominent spectral shapes such as a small
offset from systemic redshift and/or a double peak. Examining the
Ly$\alpha$ spectral profiles of $z>3$ LyC emitting LAEs will be an
interesting future endeavour.

\subsection{Rest-frame optical line ratios}

There has been accumulating observational evidence that there is a 
correlation between the [O{\sc iii}] emission line strength and LyC 
leakage. \citet{Nakajima2014} and \citet{Nakajima2016} found that
star-forming galaxies at $2<z<4$, especially LAEs, have higher
\oxyrat\ line ratios than star-forming galaxies at lower redshifts, and
LyC emitting galaxies show even higher \oxyrat\ ratios,
O32=[O{\sc iii}]$\lambda$4959,5007/[O{\sc ii}]$\lambda$3727 as high as
$>10$. \citet{Nakajima2016} argued that these high
\oxyrat\ ratios are not only due to lower metallicities but are caused
by harder ionizing radiation and/or density bounded nebulae, all of
which are properties that could lead to strong LyC
radiation. \citet{Fletcher2018} found a high LyC detection rate
($\sim$30\%) for $z=3.1$ LAEs with intense [O{\sc iii}] emission (see,
however, \citet{Naidu2018} who found no LyC emitting galaxy among
photometrically selected strong [O{\sc iii}] emitters at $z\sim3.5$).
In the lower redshift,
\citet{Izotov2016b, Izotov2016a, Izotov2018a, Izotov2018b} found LyC
emitting galaxies at $z\sim0.3$ by selecting compact star-forming
galaxies with a high O32 ratio.  
The LyC escape fraction of these galaxies are estimated to be as high as
$\sim$70\%.

\begin{figure}
 \includegraphics[width=\columnwidth]{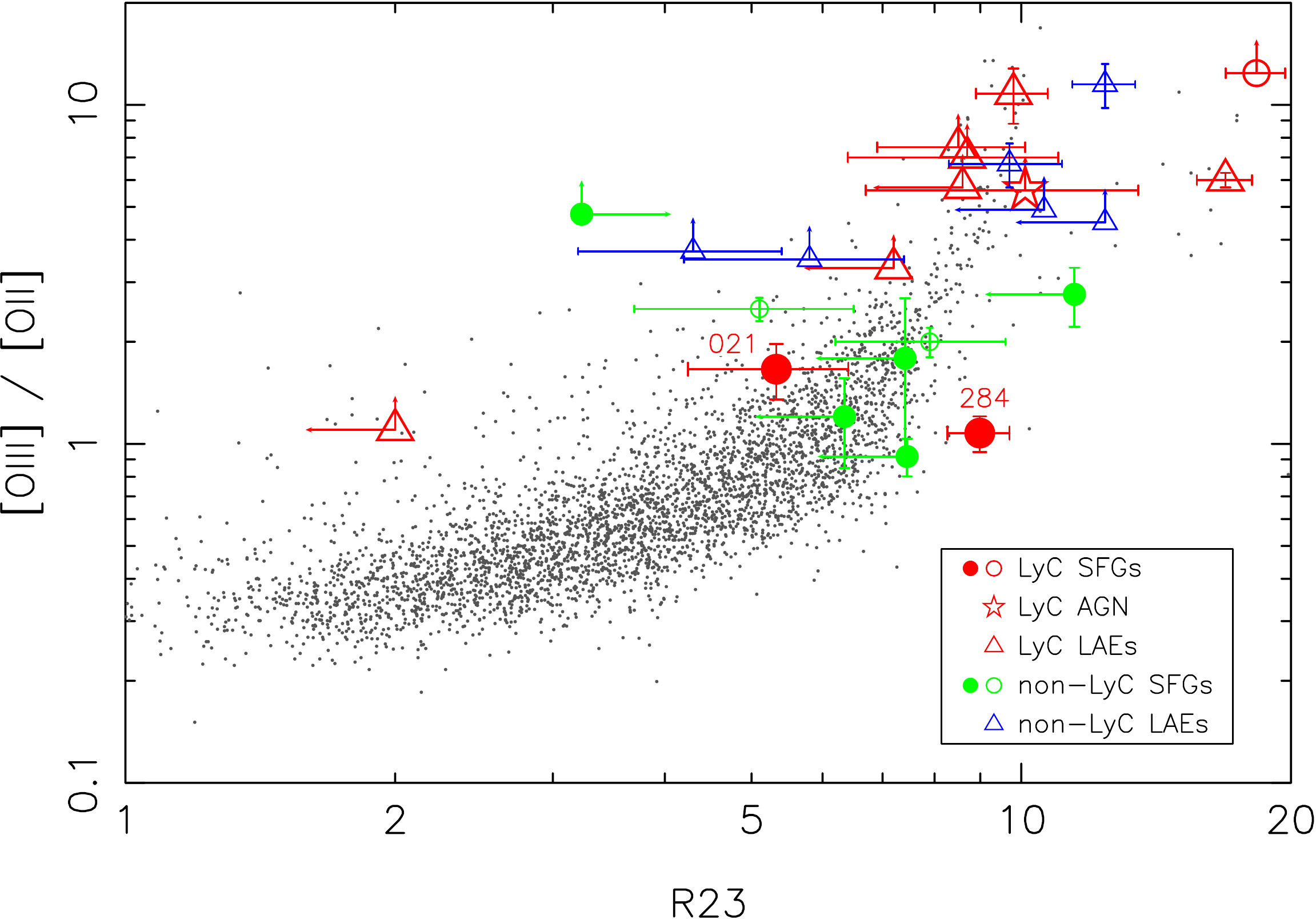}
 \caption{O32 versus R23 line ratios. Red filled circles are the two LyC
 emitting galaxy candidates among the SFGs in this study. The red open
 circle is \textit{Ion2} at $z=3.21$ \citep{deBarros2016}. Open
 triangles are $z\sim3$ LAEs with LyC detections in the literature
 \citep{Nakajima2016, Fletcher2018}. The red open star symbol is an AGN
 with id 86861 in \citet{Fletcher2018}.
 Green circles are
 $z\sim3$ SFGs/LBGs without LyC detection (filled circles are from the
 sample in this study, and open circles are from the literature), and
 open blue triangles are $z\sim3$ LAEs without LyC detection in the
 literature. Black points are low-$z$ star-forming galaxies from the
 SDSS.
 }
 \label{fig:O32_R23}
\end{figure}

Table~\ref{tab:opticallines} gives a summary of the optical emission
line measurements for a subset of the SFG sample galaxies.
In Fig.\,\ref{fig:O32_R23} we show R23 =
([O{\sc iii}]$\lambda$4959,5007+[O{\sc ii}]$\lambda$3727)/H$\beta$
and O32 for two LyC emitting candidate SFGs and five other SFGs in our
sample with [O{\sc iii}] and [O{\sc ii}] measurements, along with
values of $z\sim3$ LBGs and LAEs in the literature.
The dust attenuation of our sample galaxies is estimated by fitting
their photometric data from UV to IR compiled by \citet{Skelton2014}
with model SEDs generated using the population synthesis code P\'{E}GASE
\citep{Fioc1997} version 2 with Salpeter IMF, various metallicities and
star-formation histories.
Dust attenuation has been applied to the model SEDs assuming the
attenuation curve determined by \citet{Reddy2015}. After finding the
best-fit $E(B-V)$ value for stellar populations, attenuation for the ISM
emission lines has been estimated using an empirical relation between
the specific star-formation rates and the difference between the gas and
continuum colour excesses proposed by \citet{Reddy2015}.
Then the optical line fluxes are corrected for dust attenuation
using the attenuation curve by \citet{Cardelli1989} to derive O32 and
R23.
For comparison, we also show the distribution of SDSS star-forming
galaxies in the plot.
The locations of the two LyC candidate SFGs in this study are within the
range of the SFGs without LyC detection. This is in contrast with the
high O32 value (12.4) of \textit{Ion2} \citep{deBarros2016} which
is similar to the values of (LyC and non-LyC) LAEs.

\begin{landscape}
\begin{table}
 \centering
\caption{Rest-frame optical emission line measurements for a subset of the SFG sample galaxies$^a$.
}
\label{tab:opticallines}
\begin{tabular}{clccrrrcccrr}
\hline
 & Designation$^b$ & R.A. & Decl. & Redshift & \multicolumn{1}{c}{[O{\sc ii}]$^c$} & \multicolumn{1}{c}{H$\beta^c$} & \multicolumn{2}{c}{[O{\sc iii}]$^c$} & $E(B-V)^d$ & \multicolumn{1}{c}{O32$^e$} & \multicolumn{1}{c}{R23$^e$} \\
\cline{8-9}
 &     & \multicolumn{2}{c}{(J2000)} & &\multicolumn{1}{c}{$\lambda$3727} & &$\lambda$4959 &$\lambda$5007 & & & \\
\hline
021 & R06-D23 & 189.08048 & 62.25039 & 3.123 & $3.14\pm0.98$ & $2.87\pm0.48$ & $1.67\pm0.50$ & $8.53\pm0.62$ & 0.24 & $1.66\pm0.31$ & $5.33\pm1.08$\\
071 & S03-M35 & 189.18839 & 62.28120 & 3.235 & $3.30\pm0.81$ & $<$1.87       & $2.12\pm0.61$ & $7.18\pm0.78$ & 0.27 & $1.79\pm0.90$ & $>$7.42\\
100 & S03-M25 & 189.21167 & 62.24569 & 3.112 & $4.51\pm0.75$ & $<$1.56       & $3.59\pm0.53$ & $9.62\pm0.56$ & 0.03 & $2.76\pm0.55$ & $>$11.46\\
130 & HPS447  & 189.24792 & 62.22806 & 3.129 & $<$1.04       & $2.01\pm0.14$ & $1.06\pm0.11$ & $4.34\pm0.13$ & 0.08 & $>$4.76       & $<$3.23\\
136 & S03-M23 & 189.26126 & 62.24054 & 3.217 & $3.03\pm0.68$ & $<$1.93       & $2.79\pm0.26$ & $4.34\pm0.30$ & 0.46 & $1.20\pm0.36$ & $>$6.35\\
146 & R06-C40 & 189.27579 & 62.25272 & 3.240 & $2.43\pm0.24$ & $<$0.83       & $0.70\pm0.12$ & $2.37\pm0.18$ & 0.18 & $0.92\pm0.12$ & $>$7.46\\
284 & MOSDEF 06336& 189.27660 & 62.17246 & 3.412 & $1.86\pm0.20$ & $0.79\pm0.04$ & $1.26\pm0.11$ & $2.69\pm0.11$ & 0.24 & $1.08\pm0.13$ & $8.98\pm0.71$\\
\hline
\end{tabular}
\begin{flushleft}
\textit{Notes.}
$^a$: Measurements for object 284 (MOSDEF~06336) are done using spectra
 taken by MOSDEF.
The other values are from our MOIRCS spectroscopy.
$^b$: Origin of the sample galaxies:
021,071,100,136,146: \citet{Reddy2006},
130: \citet{Adams2011},
284: \citet{Kriek2015}.
$^c$: Line fluxes are in units of 10$^{-17}$ erg\,s$^{-1}$\,cm$^{-2}$.
Dust attenuation is not corrected.
For undetected emission lines, $3\sigma$ upper limits are given.
$^d$: Dust attenuation estimated by SED fitting (see text for details).
$^e$: O32 and R23 indices are corrected for dust attenuation using the
$E(B-V)$ value from the SED fitting.
\end{flushleft}
\end{table}
\end{landscape}

\subsection{LyC emitting AGNs}

Among the 8 AGNs in our sample, two objects are detected in \flyc,
namely, object 063 (MOSDEF 08780) and object 005 (B02-049) (see
Figs.~\ref{fig:montage_SFGAGN} and \ref{fig:montage_AGN005}). Object
063 has a complex morphology, and with detailed optical spectroscopic
observations reported by \citet{Jones2018} it was found that there is a
foreground galaxy at $z=0.512$ (their GN-UVC-2).
Among the remaining seven AGNs, while the absolute UV magnitude of one
object is relatively bright ($-25.6$), the others have
$M_{\mathrm{UV}}>-23$. The small frequency of \flyc detection (only one
among six faint AGNs) is similar to the case in the SSA22 field
reported in \citet{Micheva2017a}, and the escape fraction of $z\sim 3$
AGN is likely not to be unity \citep[see also][]{Grazian2018}.

\section{Discussion}

\subsection{Probabilities of foreground contamination}
\label{subsec:foreground}

\begin{figure}
 \includegraphics[width=\columnwidth]{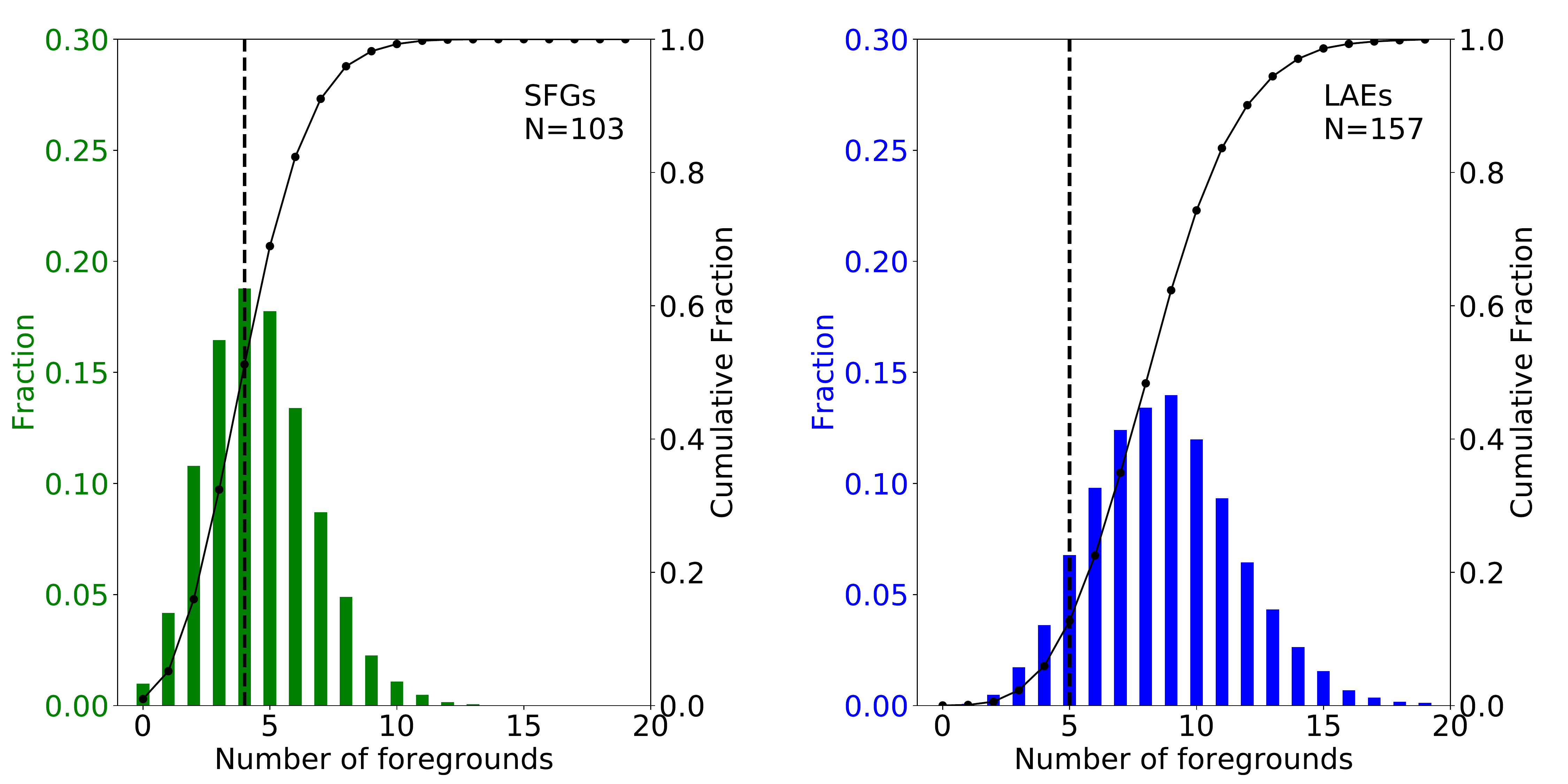}
 \caption{
 Results of the foreground contamination simulations for SFGS (left) and
 LAEs (right) in the GOODS-N. The coloured histogram shows the normalized
 distribution of the number of $\geq$3$\sigma$ detections among
 \textit{N} random apertures placed on the \flyc image.
 The black curve is the cumulative distribution of tests with the number
 of $\geq$3$\sigma$ detections equal to or smaller than \textit{n};
 i.e., the probability of having foreground contamination equal to or
 smaller than \textit{n}. 
 Horizontal dashed lines show the numbers of $\geq$3$\sigma$ detections
 in the actual GOODS-N data. See text for further details.
 }
 \label{fig:foreg_test}
\end{figure}

We estimate the number of possible contamination by chance overlap of
foreground sources in a similar way to one of the methods tested in
\citet{Micheva2017}. We randomly place \textit{N} number of $1\farcs2$
diameter apertures on the \flyc image. For SFGs \textit{N} is 103, and
the locations of the random apertures are confined to the \textit{HST}
ACS GOODS-N region, reflecting the distribution of actual SFG sample
galaxies. For LAEs \textit{N} is 157, and the entire Suprime-Cam image
area (where both \flyc and \flya images are available) is used.
We count the number of apertures with $\geq$3$\sigma$ values in the
\flyc image. We repeat this procedure 10,000 times.  
In Fig.\,\ref{fig:foreg_test} the distribution of the number of
apertures with $\geq$3$\sigma$ flux densities, as well as its cumulative 
distribution, is plotted for SFGs and LAEs.
When placing random apertures, we do not avoid the positions of objects
detected in other bands, and therefore these numbers would give
conservative estimates of the number of foreground contaminations in our
sample, likely to be an upper limit of the number of possible
contaminations.

Among the 103 sample SFGs, there are four objects with
$\geq$3$\sigma$ signal in \flyc in a $1\farcs2$ diameter aperture. One
is affected by a nearby bright source and another is shown to be
contaminated by a foreground source via the examination of its spectrum
(Section~\ref{subsec:nbphot}).
In our contamination simulation, the probability of having $\leq$3
foreground contaminations is 32.4\%, and the probability of $\geq$4
foreground contaminations is 67.6\%. Therefore, statistically there is
reasonable probability that the remaining two LyC candidates in our
SFG sample are also affected by foreground sources.
Similarly, for LAEs there are 5 objects with $\geq$3$\sigma$ signal in a 
$1\farcs2$ diameter aperture, and the simulation shows that the
probability of having $\leq$4 foreground contaminations among 157 sample
galaxies is 6.0\% (i.e., the probability of having 5 or more foreground
contamination is 94\%).

Another way to estimate probabilities of foreground contamination is to
use $U$-band number counts \citep{Vanzella2010a, Micheva2017}.
Using the $U$-band number counts based on the results by
\citet{Nonino2009} given in Table~1 of \citet{Vanzella2010a}, the number
density of objects in the magnitude range of our \flyc detected sources
(between 25.5 and 27.1 AB magnitudes) is  $\sim$190,800 deg$^{-2}$. 
If we use $1\farcs0$ aperture radius as a threshold for a chance
overlap, and assume a random distribution of objects, the probability to
have an overlap for an individual object is 4.6\%. For SFGs, the
probability to have four or more cases of foreground contamination among
103 sample galaxies is 70.7\%. For LAEs, the probability to have five or
more cases of foreground contamination among 157 sample galaxies is
85.6\%.
However, if we choose $0\farcs5$ as a threshold for a chance overlap,
which may be more appropriate considering the small spatial offsets 
for most of our candidates between the positions in \flyc and those in
rest-frame non-ionizing UV images (Table~\ref{tab:lyc_properties}),
the probability of four or more cases of foreground contamination among
103 SFGs is reduced to 3.2\% and the probability of five or more cases
of foreground contamination among 157 LAE candidates is 3.7\%.
The probabilities in these simulations are sensitive to settings such as
the aperture size and the assumed number density, and our estimates of
the probability that all of our \flyc detected sources are contaminated
by foreground sources are different for the different types of
simulations.
However, with these simulations, statistically we cannot rule out the
possibility that all detections in \flyc in the GOODS-N are affected by
foreground sources and there is no genuine LyC emitting galaxy in the
sample.

In \citet{Micheva2017} it was argued that based on a similar simulation
using the actual \flyc image in the SSA22 field, the number of
detections (21 among 159 LAEs) is difficult to explain solely by
foreground contamination. Such a different argument comes from the fact
that there appears to be a larger frequency of \flyc detections in the
SSA22 field. We discuss the differences in LyC emission between the
GOODS-N and SSA22 $z\sim 3$ galaxies in the next section.

\subsection{Comparison with the SSA22 field}
\label{sec:comparison_ssa22}

\begin{figure}
 \includegraphics[width=\columnwidth]{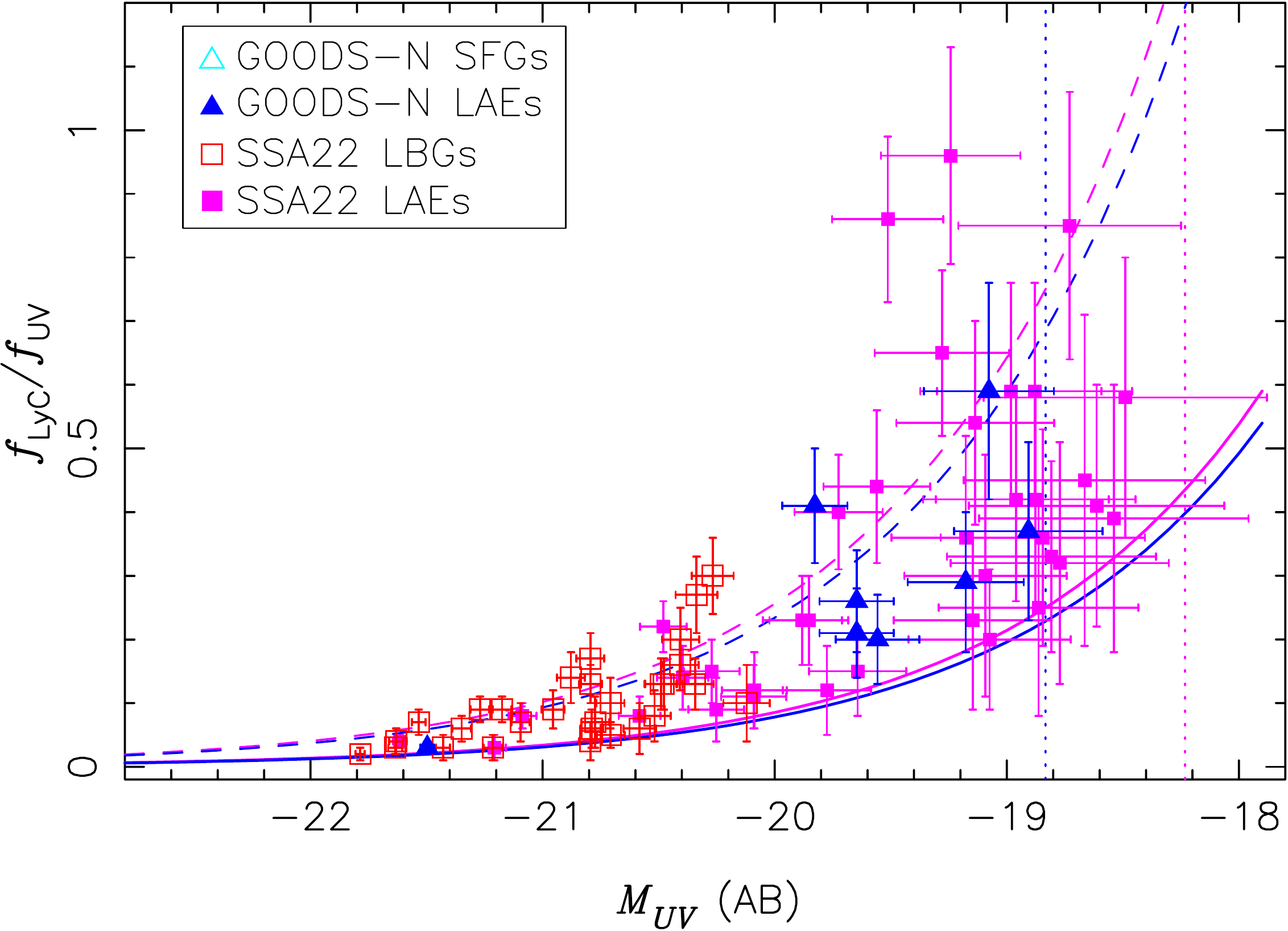}
 \caption{
 Absolute non-ionizing UV magnitudes and LyC to UV flux ratios
 for galaxies in the GOODS-N and the SSA22 fields, measured with
 $1\farcs6$ diameter apertures. Filled (open)
 triangles are LAEs (SFGs) in the GOODS-N field, and filled (open)
 squares are LAEs (LBGs) in the SSA22 field. Solid (dashed) curves
 show $1\sigma$ ($3\sigma$) detection limits for \flyc
 images. The horizontal dotted lines indicate the $3\sigma$ limiting
 magnitudes in $I_c$-band (the GOODS-N, blue) and in $i$-band (the
 SSA22, magenta) for a galaxy at $z=3.1$.
 }
 \label{fig:Muv_LyCUVrate}
\end{figure}

Because we have obtained Suprime-Cam images for the SSA22 field using
the same \flyc filter, it is possible to compare LyC emitting properties
of $z\sim3$ galaxies of the field with those in the GOODS-N field, to
examine if there are any differences in such properties depending on the
environment.
In the SSA22 field we have two types of sample galaxies. One is
LAEs selected with Suprime-Cam \flya imaging and the other is LBGs with
the drop-out selection using broad-band filter colours. Because the
selection methods of galaxies categorized as the star-forming galaxies
(SFGs) in the GOODS-N are not necessarily restricted to the drop-out
method, we have referred to this sample as SFGs. However, the SFGs in
the GOODS-N studied in this paper are broadly similar to LBGs. While the
$I$-band magnitude range of the GOODS-N SFGs is fainter than that of the
SSA22 LBGs (Fig.\,\ref{fig:Muv_LyCUVrate}), their rest-frame UV colour
distributions are almost identical.

As explained in Section \ref{subsec:nbphot}, the selection procedure we
adopted in this paper is different from those in \citet{Iwata2009} and
\citet{Micheva2017}.
In order to make a fair comparison, we constructed samples of LAEs and
LBGs in the SSA22 field with the same procedure as for the GOODS-N
sample galaxies; i.e., selecting sources with $\geq$3$\sigma$ detection
in a $1\farcs2$ diameter aperture in \flyc at the positions in
\textit{R}-band (for LBGs) and \textit{NB497} (for LAEs).
Here we use only those galaxies in the SSA22 field that have
spectroscopic redshifts.
We incorporated new spectroscopic redshifts from the SSA22 field
H{\sc i} Tomography Survey (Mawatari et al., in prep.), which is based
on optical spectroscopy with Keck / DEIMOS and compiles
spectroscopic observations in the field. The resulting numbers of sample
galaxies in the SSA22 field are 121 LBGs at $3.06<z<3.50$ and 209 
$z\sim 3.1$ \flya-selected LAEs.
There are 9 LBGs and 9 LAEs with $\geq$3$\sigma$ \flyc signal at the
position of non-ionizing UV emission detections (for LBGs) or at the
position of \flya detections (for LAEs). Among them, the number of
sources known to be contaminated by foreground sources through 
follow-up spectroscopy are 2 for LBGs and 4 for LAEs, respectively.

\begin{figure}
 \centering
 \includegraphics[width=0.9\columnwidth]{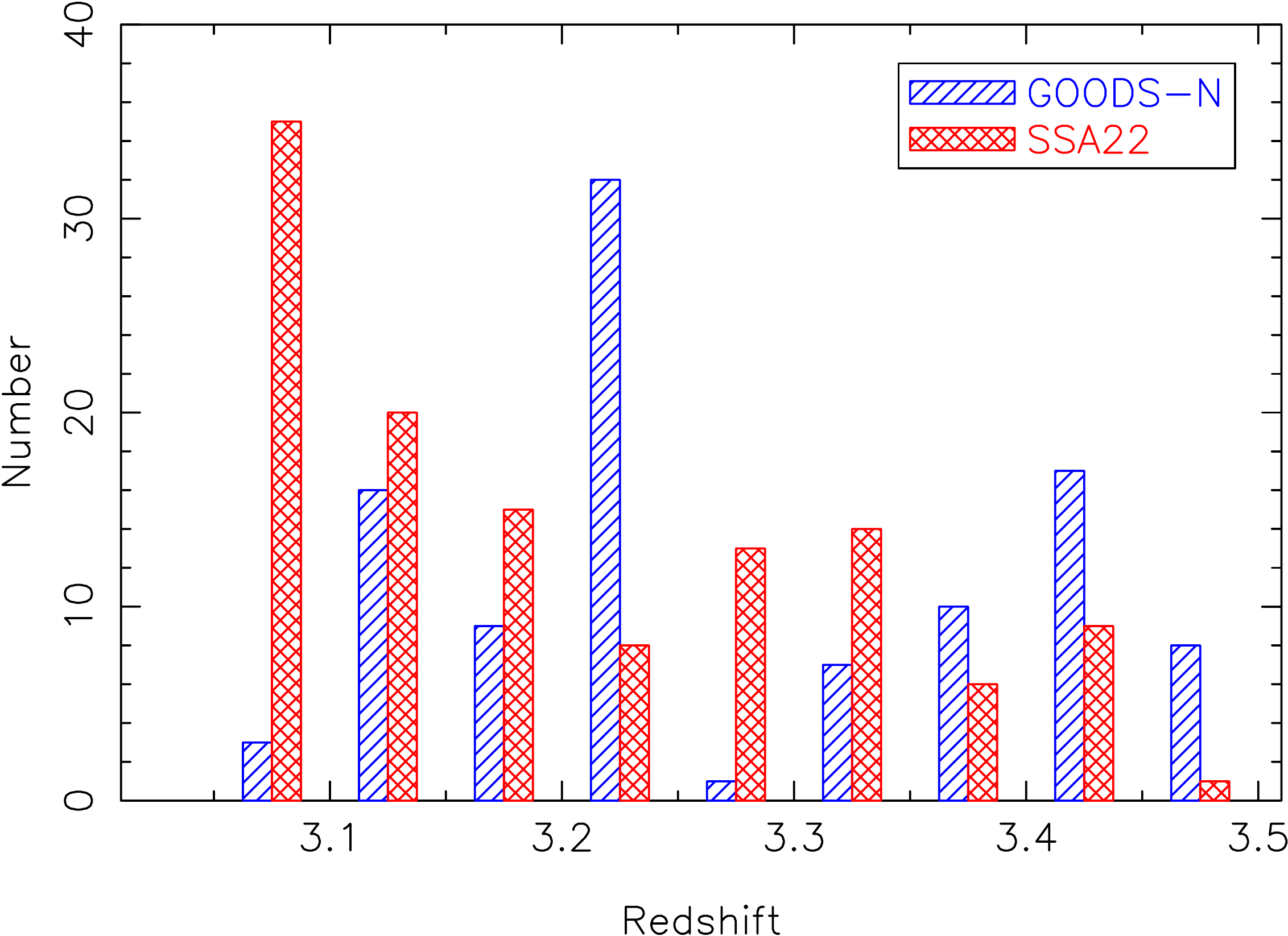}
 \caption{Comparison of redshift distributions of SFGs in the GOODS-N
 field (blue hatched) and LBGs in the SSA22 field (red cross-hatched).
 }
 \label{fig:zhist_GN_SSA22}
\end{figure}

When we compare the LyC emission from SFGs/LBGs in these two fields,
attention should be paid to the difference in redshift distribution. The
average IGM opacity against LyC rapidly increases at $z>3$ (see
Fig.\,\ref{fig:igm_trans}), and a difference in the redshift
distribution could significantly modulate the observed LyC flux
density. As shown in Fig.\,\ref{fig:imaghist} and
\ref{fig:zhist_GN_SSA22}, the distribution of SFGs in the GOODS-N peaks
at $3.20<z<3.25$ while many LBGs in the SSA22 are around $z=3.1$ where
the protocluster exists.
For LAEs we use the same \textit{NB497} filter for selection of galaxies
having strong emission around $z=3.1$ and they should reside within a
narrow redshift range of $3.06<z<3.13$ in both fields.

In Fig.\,\ref{fig:Muv_LyCUVrate} we show the absolute UV magnitudes,
derived from apparent magnitudes in $I_c$ (for the GOODS-N) or 
$i$-band (for the SSA22), and the $f_{\mathrm{LyC}}/f_{\mathrm{UV}}$
ratios (flux ratios between \flyc and $I$-bands)\footnote{In the
GOODS-N we have an $I_c$-band (effective $\lambda=7979$\AA) 
image, while in the SSA22 we have an image with $i$-band (effective
$\lambda=7683$\AA). Here we refer to both filters as `$I$-band' and do
not distinguish them, because it is expected that for our sample of
star-forming galaxies, the rest-frame UV SED is close to flat in $f_\nu$
and the flux densities in $I_c$ and $i$ will be comparable.}
for the sample galaxies in the two fields, including LAEs and LBGs /
SFGs.
Here we plot all galaxies with a \flyc flux density measurement having
more than $1\sigma$ significance in a $1\farcs6$ diameter aperture, and
detected at the $\geq$3$\sigma$ level in the $I$-band. The objects known
to be contaminated by foreground sources, as well as two LAEs (LAE 053
and 137) in the GOODS-N likely contaminated by a foreground source (see
Section~\ref{subsec:nbphot}), are removed from the plot.

We see that more galaxies have larger
$f_{\mathrm{LyC}}/f_{\mathrm{UV}}$ values among the objects fainter in
$M_{\mathrm{UV}}$, although the errors become larger in fainter
objects.
We should also note that a fainter UV magnitude boosts
$f_{\mathrm{LyC}}/f_{\mathrm{UV}}$ for objects with the same LyC flux
density. With these caveats in mind, this plot would suggest that the
galaxies with large $f_{\mathrm{LyC}}/f_{\mathrm{UV}}$ are only seen in
the fainter UV magnitude range, in both a protocluster region and in a
general field at $z\sim 3$.

It is also seen from the figure that the number of points in this
plot from the galaxies in the SSA22 are more abundant than those in the
GOODS-N field. The number of sample galaxies is different: in the
GOODS-N, the numbers of SFGs and LAEs are 103 and 158, respectively,
while for the SSA22, the numbers of LBGs and LAEs are 121 and 209,
respectively. If we consider subsamples with the same limit in
$I < 26.74$ (which is a $3\sigma$ limit for the GOODS-N, with the
corresponding limit of 27.35 in the SSA22 $i$-band), the number of 
sample galaxies are 91 SFGs and 46 LAEs in the GOODS-N, and 116 LBGs and 
112 LAEs in the SSA22. The numbers of $\geq$3$\sigma$ detections with a 
$1\farcs2$ diameter aperture in \flyc are 2 SFGs (2.2\%) and 1 LAE
(2.2\%) in the GOODS-N, while in the SSA22 they are 7 LBGs 
(6.0\%) and 5 LAEs (4.5\%). In Table~\ref{tab:gn_ssa22_stats} we
summarize these numbers, as well as the numbers of $\geq$2$\sigma$
detections.
The frequencies in the SSA22 field are always higher than those in the
GOODS-N field, although the number of significant LyC signal detections
are small for both fields.
Also, although the objects known to be contaminated by foreground
objects are eliminated, from statistics discussed in
Section~\ref{subsec:foreground}, we expect that there are still objects
contaminated by foreground objects (i.e., cases where \flyc flux
measurement does not represent LyC at $z>3$) in the plot.

\begin{table}
 \centering
 \caption{Comparison of numbers of galaxies in the GOODS-N and the SSA22
 fields.$^a$}
 \label{tab:gn_ssa22_stats}
 \begin{tabular}{lccccc}
  \hline
   & \multicolumn{2}{c}{GOODS-N} & & \multicolumn{2}{c}{SSA22} \\
  \cline{2-3} \cline{5-6}
           & SFGs & LAEs & & LBGs & LAEs \\
  \hline
$I<26.74$  &   91 &   45 & &  116 &  112 \\
$f_{\mathrm{LyC}}>3\sigma^b$ & 2 (2.2\%) & 1 (2.2\%) & &  7 (6.0\%)  & 5 (4.5\%) \\
$f_{\mathrm{LyC}}>2\sigma^b$ & 5 (5.5\%) & 3 (6.5\%) & & 13 (11.2\%) & 12 (10.7\%) \\
  \hline
 \end{tabular}
\begin{flushleft}
\textit{Note}: $^a$: Thresholds in $I$-band and \flyc are common to the
 both fields; i.e., for $I$-band a shallower $3\sigma$ limiting
 magnitude for the GOODS is used, and for \flyc the adopted
 $1\sigma$ value comes from that for the SSA22 which is shallower than
 the image in the GOODS-N.
 $^b$: The numbers of $\geq$3$\sigma$ or $\geq$2$\sigma$ detections in
 the $I<26.74$ sample galaxies.
\end{flushleft}
\end{table}

\begin{figure*}
 \includegraphics[width=\columnwidth]{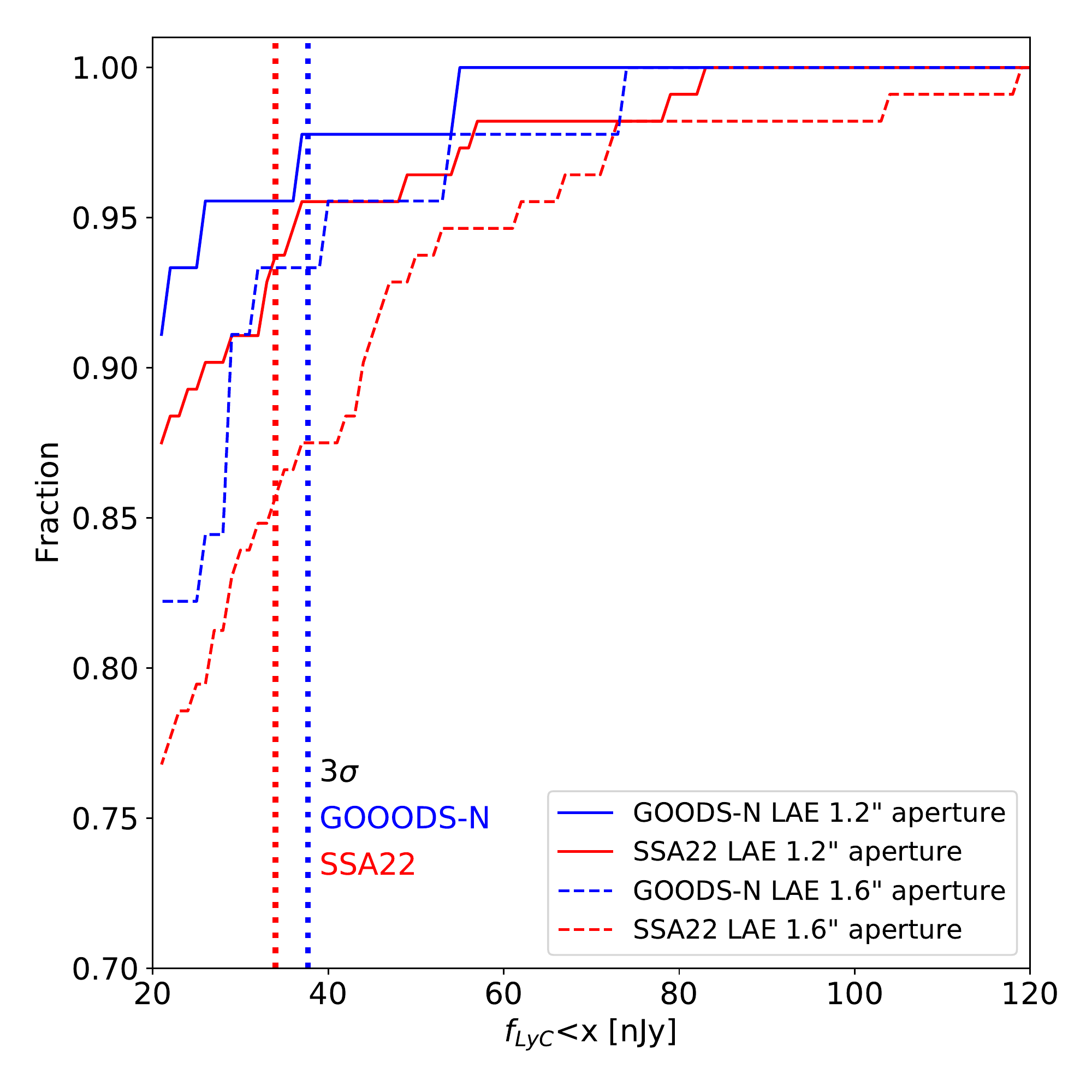}
 \includegraphics[width=\columnwidth]{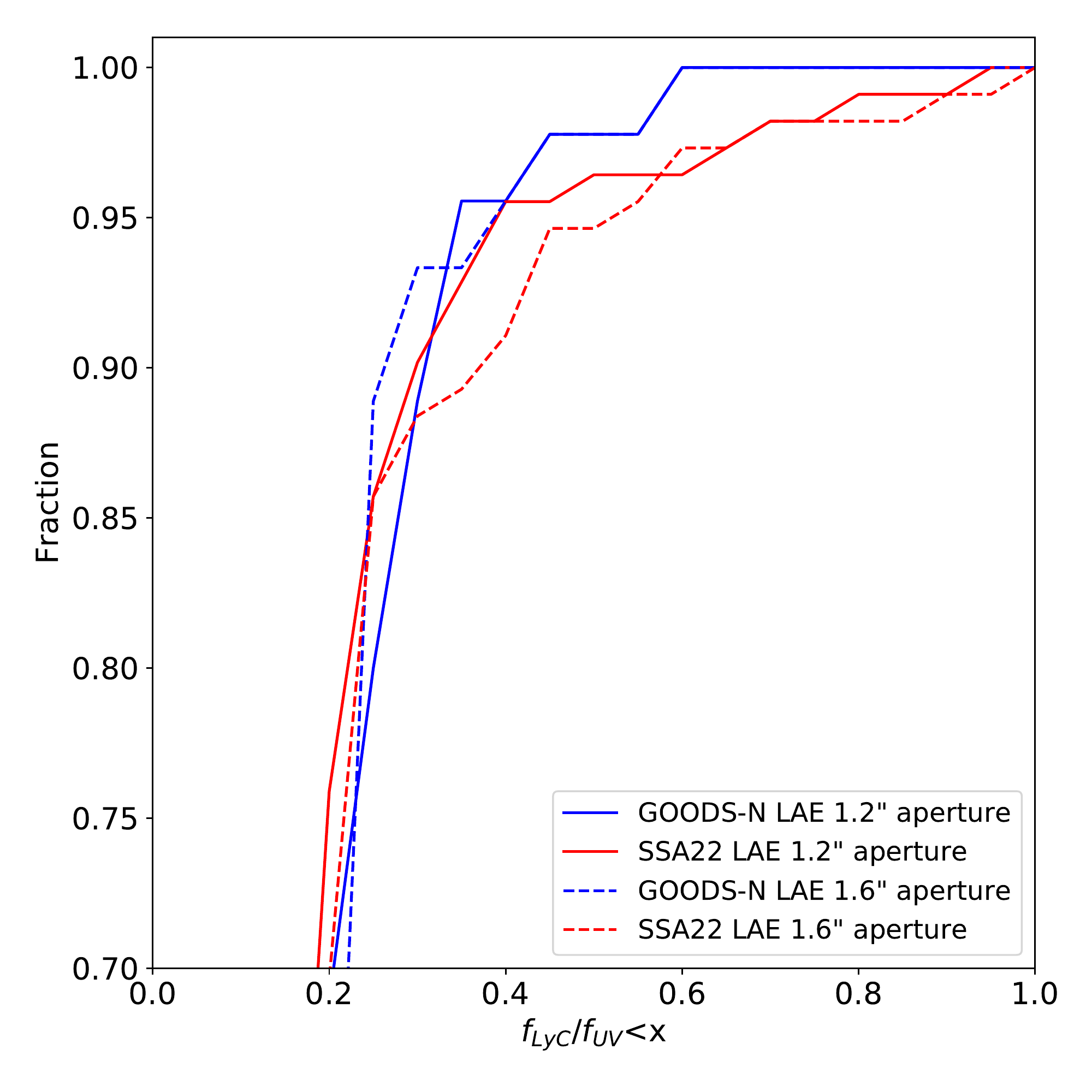}
 \caption{
\textit{(Left)}:
 cumulative distribution of LyC flux density (in $1\farcs2$ and
 $1\farcs6$ diameter apertures) for LAEs in the GOODS-N and the SSA22
 fields. Vertical dotted lines show the $3\sigma$ limits of \flyc
 images for the two fields.
 \textit{(Right)}: cumulative distribution of LyC/UV flux ratios for
 LAEs in the GOODS-N and the SSA22 fields.
 }
 \label{fig:lyc_growthcurve_LAE}
\end{figure*}

To further illustrate a possible difference between the frequency of LyC
emitting galaxies in the GOODS-N and the SSA22 fields, in
Fig.\,\ref{fig:lyc_growthcurve_LAE} the cumulative distributions of LyC
flux densities and LyC-to-UV flux ratios for LAEs
(with $I$-band magnitudes brighter than 26.74) are presented.
In this Figure, the fractions among the sample LAEs with
$f_{\mathrm{LyC}}$ (left) or $f_{\mathrm{LyC}}/f_{\mathrm{UV}}$ (right)
smaller than a specific value are plotted for both the GOODS-N and the
SSA22 fields. Here we restrict the sample to LAEs only, because for 
SFGs/LBGs it is difficult to evaluate the effect of the IGM attenuation
due to differences in the redshift distribution of the samples.
With both $1\farcs2$ and $1\farcs6$ diameter aperture measurements, the
frequency of LyC emitting LAEs appears to be higher in the SSA22 than in
the GOODS-N.
Note, however, that although the objects with spectroscopically
confirmed and likely foreground contamination have been removed, unknown
instances of foreground contamination still remaining among the sample
galaxies could affect these plots.

Because of the limited sensitivities in the LyC measurement and the
small number of detections with significant LyC emission, 
any further argument on the reality of a difference in frequency of
strong LyC emitting galaxies with these data cannot be made. However,
these comparisons suggest an interesting trend that the protocluster
field (SSA22) may contain a larger fraction of galaxies with LyC
escape. At least we can say that with the current data there is no
indication that the fraction of strong LyC emitting galaxies in a
general field (GOODS-N) is larger than in a protocluster field
(SSA22). This is counter-intuitive, because there is some observational
evidence that the amount of neutral gas in the SSA22 inter-galactic
space is enhanced compared to the cosmic average
\citep[e.g.,][]{Mawatari2017},
and one may expect that the frequency of LyC detections would therefore
be suppressed in protocluster fields. To examine the significance of the
difference in frequency of strong LyC emission between galaxies in
protocluster fields and those in general fields, we require better
sensitivity for LyC toward multiple protoclusters around $z=3$--4, as
well as a larger survey area for general fields.

\subsection{Constraints on LyC escape fraction and LyC emissivity}
\label{subsec:lyc_emissivity}

\subsubsection{LyC escape fraction}

We can put constraints on the average LyC emissivity by a stacking
analysis. First we did analyses with SFGs and LAEs in the GOODS-N
field. Because there is a statistically non-negligible possibility that
all of our sample galaxies detected in \flyc are contaminated by
foreground objects (Section~\ref{subsec:foreground}), in order to obtain
conservative constraints on \flycfuv we eliminated from the stacking all
objects with $>$3$\sigma$ detection in \flyc, as well as objects
affected by nearby bright sources.
The remaining galaxies are 99 SFGs and 151 LAEs.
We generated a calibrated \flyc image, masking all objects detected
in $I$-band other than the sample galaxies. From this image, 
10$\arcsec \times$10$\arcsec$ areas centred at each object's position in
the $I$-band (for SFGs) or in \flya (for LAEs) are extracted. Then we
normalized each cutout by the $1\farcs2$ diameter flux density of the
object in the $I$-band\footnote{The 3$\sigma$ upper limit value is used
when the object's aperture flux density is less than the value.
There is no SFG with an $I$-band flux density lower than the limit, and
the upper-limit was adopted for 65 LAEs (43\%).}, and stacked the images
to get the average value of each pixel. 
The $1\farcs2$ diameter aperture value of this stacked image gives the
averaged $(f_{\mathrm{LyC}}/f_{\mathrm{UV}})_{\mathrm{obs}}$ of the
sample.
The errors in the stacking analysis are estimated by doing the same
procedure with $N$ random ``sky'' positions in the \flyc image. In order
to take the effect of foreground source contamination into account, the
number of extracted positions $N$ is set to be the size of the entire
sample including foreground contaminations (i.e., $N=103$ for SFGs and
$N=157$ for LAEs). Images with $\geq$3$\sigma$ signal in \flyc are
excluded from stacking. This stacking with random positions is repeated
1000 times to get the standard deviation as the 1$\sigma$ error of
\flycfuv.
The measured \flycfuv\, values for the stacked images of the 99 SFGs and
the 151 LAEs both have less than $1\sigma$ errors, which means that the
errors in \flycfuv\, give the upper limits for the sample galaxies in
the GOODS-N field.
We also tried stacking analyses using sub-samples, such as galaxies with
high / low non-ionizing UV luminosities and LAEs with large Ly$\alpha$
EWs. None of the stacked images of such sub-samples showed a significant
signal in \flycfuv.

Although the frequency of galaxies with high LyC escape could be
different by environment as discussed in the previous section, the
number of sample galaxies can be increased by adding the sample galaxies
from the SSA22 field.
We selected the galaxies (SFGs/LBGs and LAEs) with redshift
$3.06 < z \leq 3.127$ from the sample in the GOODS-N and SSA22 fields. By
narrowing the redshift range, we can apply the average value of IGM
attenuation at the redshift for the ensemble of the galaxies.
There are 9 SFGs in the GOODS-N and 51 LBGs in the SSA22 with
$z\leq 3.127$ in addition to 151 (200) LAEs in the GOODS-N (SSA22).
By combining all of these galaxies, we obtain the stacked \flycfuv\,
image of 411 galaxies. Measurement with a $1\farcs2$ diameter aperture
gives \flycfuv$=0.033$ which is $\approx 1.5$ times the error estimated
by placing random apertures on the images. If we restrict the sample to
those with $\geq$3$\sigma$ detection in the $I$-band (using the same
$I<26.74$ limiting magnitude with $1\farcs6$ diameter aperture for both
fields), the number of the sample galaxies is 215. The stacking of 
these galaxies does not give significant signal ($\approx1.3\sigma$),
and from the random aperture sampling we estimate that the $3\sigma$
upper limit is \flycfuv$=0.036$. In Table~\ref{tab:stacking} the
constraints on \flycfuv\, are summarized.

The relative and absolute escape fractions of LyC are defined as
\citep{Inoue2005}:

\begin{equation}
 f_{\mathrm{esc}}^{\mathrm{rel}} =
  \frac{(L_{\mathrm{LyC}}/L_{\mathrm{UV}})_{\mathrm{out}}}{(L_{\mathrm{LyC}}/L_{\mathrm{UV}})_{\mathrm{int}}} =
  \frac{(f_{\mathrm{LyC}}/f_{\mathrm{UV}})_{\mathrm{obs}}}{(L_{\mathrm{LyC}}/L_{\mathrm{UV}})_{\mathrm{int}}} \exp(\tau_{\mathrm{LyC}}^{\mathrm{IGM}}),
\end{equation}

and

\begin{equation}
f_{\mathrm{esc}}^{\mathrm{abs}} = \frac{(L_\mathrm{LyC})_\mathrm{out}}{(L_\mathrm{LyC})_\mathrm{int}}
= f_{\mathrm{esc}}^{\mathrm{rel}} \frac{(L_\mathrm{UV})_{\mathrm{out}}}{(L_\mathrm{UV})_{\mathrm{int}}} =
f_{\mathrm{esc}}^{\mathrm{rel}} 10^{-0.4 A_\mathrm{UV}},
\end{equation}

where $(L_{\mathrm{LyC}})_{\mathrm{int}}$ and
$(L_{\mathrm{UV}})_{\mathrm{int}}$ are the intrinsic luminosity
densities of LyC and non-ionizing UV photons generated in the
star-forming regions of a galaxy, respectively, and
$(L_{\mathrm{LyC}})_{\mathrm{out}}$ and $(L_{\mathrm{UV}})_{\mathrm{out}}$
are the LyC and UV luminosity densities escaping out from the galaxy.
The ratio of observed flux densities
$(f_{\mathrm{LyC}}/f_{\mathrm{UV}})_{\mathrm{obs}}$ is modulated from
$(L_{\mathrm{LyC}}/L_{\mathrm{UV}})_{\mathrm{out}}$ by absorption of LyC
photons by intervening H{\sc i} clouds in the (circum and)
inter-galactic space.
Determining
$(L_{\mathrm{LyC}}/L_{\mathrm{UV}})_{\mathrm{int}}$
directly from observation is practically impossible for distant
galaxies, and we need to use an estimated value from e.g., population
synthesis models. Predicted values could greatly vary, depending on
various factors such as the stellar evolution models, initial mass
function, and age of the stellar populations.
Here we use
$(L_{\mathrm{LyC}}/L_{\mathrm{UV}})_{\mathrm{int}}=0.3$ as a fiducial
value, which is close to that adopted by previous studies on $z\sim3$
galaxies \citep[e.g.,][]{Iwata2009, Steidel2018}.
For the dust attenuation of non-ionizing UV photons within galaxies, we
adopt an average value of $A_{\mathrm{UV}}=1.67$, which is derived by
\citet{Micheva2017} for LAEs in the SSA22 from the observed $V-i$
colours.
The distribution of $V-I$ colours of the galaxies in the GOODS-N field
are very similar to that of the galaxies in the SSA22 field.

The mean IGM transmission from the analytic formula by
\citet{Inoue2014} is
$\exp(-\tau)_{\mathrm{LyC}}^{\mathrm{IGM}} =0.406$ at $z=3.1$ with the 
\flyc filter. If we use the IGM+CGM transmission model for the
wavelength range $880 \leq \lambda_0 \leq 910$\AA\ by
\citet{Steidel2018}, it is 0.352.
Table~\ref{tab:stacking} summarizes the constraints on
$f_{\mathrm{esc}}^{\mathrm{rel}}$ and
$f_{\mathrm{esc}}^{\mathrm{abs}}$ using the mean IGM value at $z=3.1$
with the \citet{Inoue2014} model and those on
$f_{\mathrm{esc}}^{\mathrm{abs}}$
using the IGM+CGM model by \citet{Steidel2018}.
Note that the \flyc filter traces the wavelength range $860 \lesssim
\lambda_0 \lesssim 900$\AA\ for a source at $z=3.1$, and the IGM+CGM
model by \citet{Steidel2018} would actually predict slightly smaller
value for the \flyc filter than 0.352 which is used in
Table~\ref{tab:stacking}. The IGM-only model in \citet{Steidel2018},
which should be similar to the analytic model by \citet{Inoue2014},
gives the IGM transmission for the wavelength range 
$880 \leq \lambda_0 \leq 910$\AA\ to be 0.435, which is slightly larger
than the value 0.406 for the \flyc filter with the \citet{Inoue2014}
model.

\begin{table*}
 \centering
 \caption{Constraints from stacking analyses.}
 \label{tab:stacking}
 \begin{tabular}{lrrccc}
  \hline
  Sample populations & Number & (\flycfuv)$_\mathrm{obs}^a$ & ($f_{\mathrm{esc}}^{\mathrm{rel}}$)$_{\mathrm{IGM}}^b$ & ($f_{\mathrm{esc}}^{\mathrm{abs}}$)$_{\mathrm{IGM}}^b$ & ($f_{\mathrm{esc}}^{\mathrm{abs}}$)$_{\mathrm{IGM+CGM}}^c$ \\
  \hline
  GOODS-N SFGs ($3.06<z<3.5$) & 99  & $<0.036$ & -- & -- & --\\
  GOODS-N $z=3.1$ LAEs        & 151 & $<0.099$ & $<0.81$ & $<0.17$  & $<0.20$ \\
  SSA22+GOODS-N $z=3.1^d$     & 411 & $<0.067$ & $<0.55$ & $<0.12$  & $<0.14$ \\
  SSA22+GOODS-N $z=3.1$, $I<26.74^d$ & 215 & $<0.036$ & $<0.29$ & $<0.063$ & $<0.073$\\
  \hline
 \end{tabular}
\begin{flushleft}
\textit{Note}:
 $^a$: 3$\sigma$ upper limits.
 $^b$: using the IGM model by \citet{Inoue2014}.
 $^c$: using the IGM+CGM model by \citet{Steidel2018}.
 $^d$: The sample includes SFGs in the GOODS-N
 and LBGs in the SSA22 with $z_{\mathrm{spec}} \leq 3.127$,
 and \flya selected LAEs from both fields.
\end{flushleft}
\end{table*}

These constraints based on the large sample of galaxies in the two
independent fields with the same narrow-band filter give a stringent
upper limit on the average LyC escape fraction of $z\sim 3.1$ galaxies.
The average escape fraction of star-forming galaxies at
$z\sim3.1$ with $I<26.74$ ($M_{\mathrm{UV}}<-18.8$ $\sim$0.1$L^\ast$,
where $L^\ast$ of $z\sim 3$ LBGs is $\approx -21.0$
\citep[e.g.,][]{Sawicki2006, Reddy2009}) is less than 8\% with a
correction for attenuation by intervening H{\sc i} clouds using both the
IGM model by \citet{Inoue2014} and the IGM+CGM model by
\citet{Steidel2018} (bottom line of Table~\ref{tab:stacking}).

\subsubsection{Comparison with previous studies}

Although the distribution of \flycfuv\ versus the absolute UV
magnitude for individual sources (Fig.~\ref{fig:Muv_LyCUVrate})
suggests a higher frequency of LyC escape in galaxies faint in UV
magnitudes, the average escape fraction of UV-faint star-forming
galaxies does not show a rise from the
$f_{\mathrm{esc}}^{\mathrm{abs}}$ value for the brighter galaxies in the
same redshift range; in \citet{Steidel2018}
$f_{\mathrm{esc}}^{\mathrm{abs}} = 0.09\pm0.01$ was obtained from 124
LBGs at $2.7 < z < 3.6$ through fitting their composite spectrum. 
Although the UV absolute magnitude range of their sample is
$-22.1 \leq M_{\mathrm{UV}} \leq -19.5$, the majority of the sample has
absolute magnitude $<-20.5$, about 2 magnitudes brighter than the UV
magnitude limit of the present sample.

\citet{Grazian2017} used deep $U$-band imaging data in multiple fields
to search for LyC emitting galaxies at $3.27 \leq z \leq 3.40$. They
stacked 69 SFGs in order to constrain the LyC escape fraction, and
obtained $f_{\mathrm{esc}}^{\mathrm{rel}} \leq 0.017$ (1$\sigma$).
While their sample contains some UV-faint SFGs down to
$M_{\mathrm UV} = -19$, most of their sample galaxies have UV
luminosities similar to the brighter part of the sample galaxies in the
present study, and the median absolute UV magnitude of their sample
galaxies is $-20.9$, one magnitude brighter than the median of our
sample galaxies ($-19.9$). Our constraints on the LyC
escape fraction in fainter galaxies are broadly consistent with the
results by \citet{Grazian2017}.
These authors discussed the UV luminosity dependence of the LyC escape
fraction, but could not draw a firm conclusion due to a weaker
constraint on the LyC escape fraction for UV faint galaxies compared to
bright ones. The situation is the same in the present study.

\citet{Nestor2013} and \citet{Mostardi2013} used Keck / LRIS narrow-band
imaging to select UV-faint LAEs in the protocluster regions at
$z\sim 3.1$ (the SSA22 field) and at $z\sim 2.85$, respectively, and
used different narrow-band filters to examine LyC from their sample
galaxies. Although their search field areas are smaller than ours, both
their methods and the UV luminosity distribution of their LAE samples
are similar to ours.
The constraints on the LyC escape fraction are
$f_{\mathrm{esc}}^{\mathrm{abs}} \sim$10\%--30\% in \citet{Nestor2013},
based on 91 LAEs, and $f_{\mathrm{esc}}^{\mathrm{abs}} \sim$5\%--15\% 
in \citet{Mostardi2013}, based on 91 LAEs. Our results are consistent
with their upper limits.

\subsubsection{A constraint on the LyC emissivity}

We can calculate an upper limit of the contribution of $z\sim 3$ SFGs to
the volume-averaged ionizing radiation emissivity, by integrating the
LyC luminosity over a range of UV luminosities:

\begin{equation}
\epsilon_{\mathrm{LyC}} = \int
 (f_{\mathrm{LyC}}/f_{\mathrm{UV}})_{\mathrm{out}} L_{\mathrm{UV}} \phi d L,
\label{eq_emissivity}
\end{equation}
where $L_{\mathrm{UV}}$ is the non-ionizing UV luminosity and $\phi$ is
the number density of the galaxies in the luminosity range (luminosity
function). From the stacking analysis of the sample galaxies with
$z=3.1$ and $I<26.74$
(Table~\ref{tab:stacking}) and the mean IGM transmission at $z=3.1$
(0.406 from \citet{Inoue2014}), we adopt the 3$\sigma$ upper limit of 
$(f_{\mathrm{LyC}}/f_{\mathrm{UV}})_{\mathrm{out}}
 = (f_{\mathrm{LyC}}/f_{\mathrm{UV}})_{\mathrm{obs}} \exp(\tau_{\mathrm{LyC}}^{\mathrm{IGM}}) < 0.088$.
(If we use the IGM+CGM model by \citet{Steidel2018} instead of the IGM
model by \citet{Inoue2014} the upper limit is 0.10, a factor of 1.15
higher.)
Using the UV luminosity function of $z\sim 3$ LBGs by 
\citet{Reddy2009}, from Equation~\ref{eq_emissivity} we obtain 
$1.95 \times 10^{25}$ erg s$^{-1}$ Hz$^{-1}$ Mpc$^{-3}$ as the 3$\sigma$ 
upper limit on the ionizing emissivity from galaxies with
$M_\mathrm{UV}<-18.84$.
For comparison, \citet{Becker2013} inferred $\epsilon_\mathrm{LyC}$ at
$2<z<5$ from QSO Ly$\alpha$ forest observations, and their estimate is
$2.86-21.5 \times 10^{24}$ erg s$^{-1}$ Hz$^{-1}$ Mpc$^{-3}$ at $z=3.2$.
Our upper limit on the contribution to the ionizing radiation emissivity
by SFGs (including UV-faint LAEs) does not exclude the possibility that
SFGs are the dominant source of the volume-averaged ionizing radiation
at $z\sim 3$, and is consistent with previous studies
\citep[e.g.,][]{Nestor2013, Steidel2018}.

Current constraints from the UV luminosity function of SFGs at $z>6$
require that $f_{\mathrm{esc}}^{\mathrm{abs}} \gtrsim$10\%
\citep[e.g.,][]{Bouwens2016}.
If star-forming galaxies at $z>6$ have similar low LyC escape fraction,
it may be difficult to attribute to star-forming galaxies all the
ionizing photons responsible for cosmic reionization. We may need either
a change in the physical properties of typical star-forming regions in
galaxies to enhance LyC escape from $z\sim 3$ to $z>6$
\citep[e.g.,][]{Nakajima2016} or a major contribution from other sources
of ionizing radiation such as AGNs or populations not yet observed at
$z\sim3$.

\section{Conclusions}

We conducted a narrow-band imaging search for LyC emission among GOODS-N
field galaxies at $z>3.06$. Our main findings are summarized as follows:

\begin{itemize}
 \item Among 103 star-forming galaxies (SFGs) at $3.06<z<3.5$ we found
       two candidates of LyC emitting galaxies.
       One of them has a peculiar spectral energy distribution, and
       there is a possibility of contamination by an unresolved
       foreground source. Near-infrared spectroscopy of the other
       candidate indicates a moderately elevated 
       [O{\sc iii}]/[O{\sc ii}] ratio, which is within the range of
       other LyC emitting galaxies. 
 \item Among 157 $z=3.1$ LAE candidates, there are five detections in
       \flyc, and three among them are possibly
       contaminated by foreground sources.
 \item Statistically we cannot rule out a possibility that all of the
       galaxies detection in \flyc are contaminated by foreground
       sources.
 \item A comparison between the frequency of LyC flux and LyC-to-UV flux
       ratios in the GOODS-N and the SSA22 field sample galaxies suggests
       that there is no enhancement of LyC escape in the general field
       compared to the protocluster field at $z=3.1$.
 \item By combining the sample galaxies in the GOODS-N and the SSA22
       fields, we put a stringent upper limit $<$8\% (3$\sigma$) on the
       LyC escape fraction of the $M_\mathrm{UV}<-18.8$ galaxies.
\end{itemize}

In order to further explore LyC from high-redshift galaxies via direct
observations, more sensitive observations with wider survey areas are
needed, which would yield larger number of individual LyC emitting
galaxy candidates, put more stringent constraints on the average LyC
escape fraction, and enable us to investigate any differences depending
on the galaxy environments. The on-going strategic survey with Subaru /
Hyper Suprime-Cam \citep{Aihara2018}, the associated narrow-band survey
CHORUS (Inoue~et al. in prep.) and the $U$-band survey with CFHT CLAUDS
(Sawicki~et al. in prep.) will provide such opportunities.

\section*{Acknowledgments}

The authors thank staff members of the Subaru Telescope and the Subaru
Mitaka Office for their support, and the anonymous referee for careful
reading of the manuscript and constructive suggestions.
II acknowledges Dr. Naveen Reddy and the MOSDEF team for providing
near-infrared spectra of an object used in this study.
He also thanks Dr. Chuck Steidel for providing us
with a Keck/LRIS spectrum of a galaxy, and Dr. Marcin Sawicki for
arranging a comfortable work environment for him during the analyses of
the present research.

This work was supported by JSPS KAKENHI Grant Numbers 18740114,
24244018, and 17H0114.

Some of the archived data collected at the Subaru Telescope
are obtained from SMOKA, which is operated by the Astronomy Data
Center, National Astronomical Observatory of Japan, and
STARS (Subaru Telescope Archive System) operated by the Subaru
Telescope.
Some of the data presented in this paper were obtained from the
Mikulski Archive for Space Telescopes (MAST). STScI is operated by the
Association of Universities for Research in Astronomy, Inc., under NASA
contract NAS5-26555.
This research partly uses the database of the Sloan Digital Sky Survey
(SDSS). Funding for the SDSS and SDSS-II has been provided by the Alfred
P. Sloan Foundation, the Participating Institutions, the National
Science Foundation, the U.S. Department of Energy, the National
Aeronautics and Space Administration, the Japanese Monbukagakusho, the
Max Planck Society, and the Higher Education Funding Council for
England. The SDSS Web Site is http://www.sdss.org/.

IRAF is distributed by the National Optical Astronomy Observatories,
which are operated by the Association of Universities for Research in
Astronomy, Inc., under cooperative agreement with the National Science
Foundation.
This research made use of Astropy, a community-developed core Python
package for Astronomy.

We wish to express our gratitude to the indigenous Hawaiian community
for their understanding of the significant role of the summit of
Maunakea in astronomical research.




\bibliographystyle{mnras}
\bibliography{GNLyC}



\bsp	
\label{lastpage}
\end{document}